\documentstyle[11pt,epsf]{article}

\newcommand{\be}{\begin{equation}}
\newcommand{\ee}{\end{equation}}
\newcommand{\bea}{\begin{eqnarray}}
\newcommand{\eea}{\end{eqnarray}}
\date{}
\renewcommand{\theequation}{\arabic{section}.\arabic{equation}}

\def\deli{\partial_{\kappa}}
\def\delj{\partial_{\lambda}}
\def\delk{\partial_{\mu}}
\def\delij{\partial_{\kappa\lambda}}

\def\deljk{\partial_{\lambda\mu}}
\def\delki{\partial_{\mu\kappa}}

\def\delijk{\partial_{\kappa\lambda\mu}}

\def\delijkl{\partial_{\kappa\lambda\mu\nu}}

\begin{document}
\begin{titlepage}
\begin{flushright}
HD--THEP--94--13\\
\end{flushright}
\vspace{1.5cm}
\begin{center}
{\bf\LARGE CRITICAL BUBBLES AND FLUCTUATIONS}\\
\vspace{.5cm}
{\bf\LARGE AT THE ELECTROWEAK PHASE TRANSITION}\\
\vspace{1cm}
Jochen Kripfganz$^\dagger$\footnote{supported by Deutsche 
Forschungsgemeinschaft}\\
\vspace{.3cm}
Andreas Laser$^\dagger$\\
\vspace{.3cm}
Michael G.~Schmidt$^{\dagger*}$\footnote{on sabbatical leave\\
e-mail addresses:\\
\begin{tabular}{rl} 
J.~Kripfganz & dj8@vm.urz.uni-heidelberg.de\\
A.~Laser & t82@ix.urz.uni-heidelberg.de\\
M.G.~Schmidt & k22@vm.urz.uni-heidelberg.de
\end{tabular}}\\
\vspace{1cm}
$^\dagger$Institut  f\"ur Theoretische Physik\\
Universit\"at Heidelberg\\
Philosophenweg 16\\
D-69120 Heidelberg, FRG\\
\vspace{0.5cm}
$^*$Theoretical Physics Division\\
CERN\\
CH-1211 Geneva 23, Switzerland\\
\vspace{1.5cm}
{\bf Abstract}\\
\end{center}
We discuss the critical bubbles of the electroweak phase transition
using an effective high-temperature 3-dimensional action for the 
Higgs field $\varphi$. The separate integration of gauge and 
Goldstone boson degrees of freedom is conveniently described in the 
't~Hooft-Feynman covariant background gauge. The effective dimensionless
gauge coupling $g_3(T)^2$ in the broken phase is well behaved 
throughout the phase transition. However, the behavior of the one-loop 
$Z(\varphi)$ factors of the Higgs and gauge kinetic terms signalizes 
the breakdown of the derivative expansion and of the perturbative expansion 
for a range of small $\varphi$ values increasing with the Higgs mass 
$m_H$. Taking a functional $S_z[\varphi]$ with constant $Z(\varphi)=z$
instead of the full non-local effective action in some neighborhood of 
the saddlepoint we are calculating the critical bubbles for several 
temperatures. The fluctuation determinant is calculated to high
accuracy using a variant of the heat kernel method. It gives a strong
suppression of the transition rate compared to previous estimates.
\end{titlepage}

\newpage
\section{Introduction}
There are strong indications \cite{DiLi} -- \cite{KaRuSha} that the 
electroweak standard theory 
predicts a first-order phase transition at high temperatures corresponding
to the electroweak scale. The critical bubbles of the transition are 
solutions of the static electroweak
semiclassical equations of motion. The transition probability can be 
calculated using Langer's theory \cite{Langer}. The exact production rate
is important for the timing of the transition and the determination of the
corresponding temperature. The baryon asymmetry of the universe
may be generated through bubble expansion because the three Sakharov criteria
-- C/CP violation, baryon number violation and nonequilibrium -- are
fulfilled. 

In a proper treatment of the electroweak phase transition
the coarse-grained action constructed
consistently for the particular problem and the
size scale involved should be used. If there 
are different mass scales one can integrate out first the more massive fields
and keep the light fields in an effective action relevant for the phase transition. 
At high temperature non-static Matsubara modes may be considered as heavy
fields. Integrating out these modes leads to a 3-dimensional effective 
theory with symmetry restoration at high temperature. 
The related phase transition is predicted to be second-order instead 
of first-order, however. It becomes first-order due to contributions from
static modes. So some of the light modes have to be
integrated out as well in calculating the effective action.
It is well known that
integrating out the non-static modes using the high temperature expansions
and then integrating out the static gauge boson fields one has already
an effective potential leading to a first-order phase transition.
In this spirit it is very natural to integrate out the static Goldstone
modes as well. This is most conveniently done in the 't~Hooft-Feynman
covariant background gauge. 
With the pure real scalar background of the bubble configurations 
one ends up with a 3-dimensional effective Higgs action whose fluctuations 
still have to be considered. 

If a small coupling parameter - maybe after some redefinitions - can 
be identified, the one-loop perturbative expansion for the effective action
will be a good approximation. A carefully introduced gauge coupling $g_3(T)^2$
in the broken phase is well behaved throughout the phase transition, and 
indeed it is not very large for $m_H\leq m_W$.

The $Z_H(\varphi)$ factor of the Higgs kinetic term will be calculated explicitly
in the 't~Hooft-Feynman covariant background gauge in one-loop order. It differs
considerably from that in the Landau gauge. Due to the fact that 
$Z_H(\varphi)$ is negative in some $\varphi$ range, it is not sensible to include it in this
form in the effective action. We will argue that the derivative expansion
breaks down in this range.

We also inspect the one-loop $Z_{\rm gauge}(\varphi)$ prefactor of the kinetic gauge term.
It plays no direct role in the bubble action. However, $g_3(T)^2/Z_{\rm gauge}(\varphi)$ is 
the effective gauge coupling at the constant scale $\varphi$. 
It blows up for $Z_{\rm gauge}\to0$ and this happens already at rather large values of
$\varphi \; (>0.4)$ for $m_H>\frac{1}{2}m_W$.
Thus except for very small $m_H$ the $\varphi$-range relevant for the bubble 
configuration may be largely nonperturbative in the effective gauge coupling.
Integration of the (enlarged) gauge degrees of freedom to some (one) loop order
can in this range at best be suggestive for the form of the effective Higgs 
action. 

We will then consider a Higgs effective action with the known one-loop
effective potential and a kinetic term with arbitrary constant $Z$ factor.
This allows us to come to our main subject, the discussion of radiative
corrections in the heat kernel method.
Since there is no background gauge field for the bubble
configuration, the heat kernel expansion already exists to very high order
(and can easily be extended using new methods \cite{SchmSchu, FliSchmSchu}).
Thus we can arrive at a very precise treatment of the fluctuations.

It turns out that the static prefactor which is essentially the 
fluctuation determinant gives a strong suppression
of the nucleation rate. Its logarithm may be interpreted as the one-loop
correction to the effective action of the critical bubble. The comparison 
between the two values decides on the applicability of the
nucleation theory. 

Section 2 contains a discussion of the 3-dimensional effective high temperature
Higgs action obtained from one-loop integration of all the other fields
in the  't~Hooft-Feynman background gauge and a critical inspection of
derivative expansion and of perturbation theory. In section 3 we first shortly
review the nucleation rate based on Langer's  theory. We discuss the critical
bubbles obtained with the modification mentioned above. In the following we 
consider the heat 
kernel method for calculating the fluctuation determinant and develop a
particular method to treat zero/instable modes. We present our high
accuracy results for the fluctuation prefactor in the transition rate. 
Section 4 gives our conclusion.
Appendix A contains the calculation of $Z$ factors, Appendix B some
generalization of the thin-wall bubble solution. Appendix C gives the
first six operators in the heat kernel expansion.

\section{The Effective Action}
\setcounter{equation}{0}

The critical bubble solutions describing
the first-order phase transition of the electroweak theory are
 pure Higgs field
configurations. They are not solutions of the original fundamental field
equations of  the electroweak theory, but correspond to an 
effective action where (part of)
the other field degrees of freedom have already been integrated 
out. It
is the aim of the `exact renormalization group approach' 
to derive such an action well adapted to the size of the bubbles.
In the case of a gauge theory this demanding program is just being
developed \cite{ReuWe, Ellwanger}. Still a simpler way to discuss 
critical bubbles is to generate new terms of the Higgs effective action by 
using low-order perturbation theory, starting from the fundamental Lagrangian.
However in a rigorous treatment this requires the identification of appropriate
expansion parameters (which may not be possible).

\subsection{The high-temperature effective action}

In the case of the  electroweak phase transition 
it is appropriate to perform a high temperature expansion.
The expansion parameter is $m(T)/T$. The guiding principle is to integrate 
out the heavy field degrees of freedom to get an effective theory of the light
fields.

In a first step it is possible to integrate out the non-static Matsubara-frequencies
which gain masses proportional to the temperature $T$ ($2n \pi T$ with 
$n \neq 0$ for bosons and $(2n+1) \pi T$ for fermions). The remaining effective
theory is purely bosonic and 3-dimensional. As argued in ref.~\cite{Landsman}
the high-temperature dimensional reduction has shortcomings in higher orders 
of perturbation theory. 

In a second step the longitudinal component $A_0$ of the gauge field 
is integrated out. It develops a Debye-mass proportional to $gT$. 

In a third step we rescale the coordinates and fields 
\be \label{rescale}
\vec{x} \rightarrow \frac{\vec{x}}{gv}, \qquad   \Phi \rightarrow v\Phi,
 \qquad   A \rightarrow v A
\ee
where the scale $v$ is left open for the moment. The remaining 
high-temperature effective action can be written, in the limit of vanishing 
electroweak mixing angle, as
\be\label{Sht}
S_{\rm ht} = \frac{1}{g_3(T)^2}  \int d^3x \left[ \frac{1}{4} F_{ij}^a F_{ij}^a 
 + (D_i\Phi)^\dagger (D_i\Phi) + 
 V_{\rm ht}(\Phi^\dagger\Phi) \right]
\ee
with other contributions vanishing 
powerlike at high temperature, 
e.g.~a $(\Phi^\dagger \Phi)^3 /T^2$ term. They have been
discussed to be unimportant \cite{JaKaPa}.

The effective 3-dimensional gauge coupling is
defined as 
\be
g_3(T)^2 = \frac{g T}{v} \quad.
\ee
The gauge coupling $g$ has been scaled out of the covariant derivative 
and the field strength tensor.
The high-temperature effective potential is 
\be
V_{\rm ht}(\Phi^\dagger\Phi) = 
   \frac{\lambda_T}{g^2}\left( (\Phi^\dagger\Phi)^2 - 
   \left(\frac{v_0(T)}{v}\right)^2\Phi^\dagger\Phi \right) \quad.
\ee
$\frac{v_0(T)^2}{v^2}$ is the asymmetric minimum 
\be
v_0(T)^2 = \frac{2}{\lambda_T} \left(T_0^2 - T^2 \right) D \quad.
\ee
It is negative for $T>T_0$. At these temperatures the global minimum of 
$V_{\rm ht}(\Phi^\dagger \Phi)$ is the symmetric one at 
$\Phi^\dagger \Phi = 0$. At $T_0$ it
moves continuously to finite values. Therefore this potential predicts a 
second-order phase transition. 

The constants are determined by the parameters of the standard model. They
can be calculated from the zero temperature masses $\tilde{m}$ and the zero 
temperature vacuum expectation value of the scalar field $\tilde{v} = 246$ GeV 
\begin{eqnarray}
\tilde{m}^2_W &=& \frac{1}{4} g^2 \tilde{v}^2, \qquad  
\tilde{m}^2_H = 2\lambda \tilde{v}^2 \nonumber\\
T_0^2 &=& \frac{ \tilde{m}_H^2 - 8\tilde{v}^2 B }{4D} \nonumber\\
D &=& \frac{1}{ 8\tilde{v}^2}(3\tilde{m}_W^2 + 2\tilde{m}_t^2) \nonumber\\
B&=&\frac{3}{64 \pi^2 \tilde{v}^4}(3\tilde{m}_W^4 - 4\tilde{m}_t^4)  \quad.
\end{eqnarray}
The temperature dependent quartic coupling is 
\begin{eqnarray}
\lambda_T &=& \lambda - \frac{3}{16 \pi^2 \tilde{v}^4} 
  \left(3\tilde{m}_W^4 \log\frac{\tilde{m}_W^2}{a_B T^2} \;-\; 
        4\tilde{m}_t^4 \log\frac{\tilde{m}_t^2}{a_F T^2} \right)
\nonumber\\
&& \ln(a_B) = 3.91 , \qquad \ln(a_F) = 1.14 \quad.
\end{eqnarray}
The theory described by eq.~(\ref{Sht}) is nothing but the 3-dimensional 
$SU(2)$ Higgs model; it has 13 field degrees of freedom; 9 from the gauge 
field and 4 from the scalar field. 

\subsection{Background field and fluctuations}

The high-temperature potential in eq.~(\ref{Sht}) corresponds to a 
second-order phase transition. From lattice calculation 
\cite{BuIlKriSchi, KaRuSha}, however,
and from the full one-loop effective potential \cite{DiLi} -- \cite{JaKaPa}
one expects the electroweak phase transition to be 
first-order. This will be manifest by integrating out further degrees of freedom. 

In most cases, a first-order phase transition is initiated
by the formation of critical bubbles. The electroweak critical bubbles 
are pure real Higgs field
configurations. We therefore divide the 13 remaining fields into the real
background field $\varphi$ and into fluctuations in the following way 
\begin{eqnarray}
A_i^a &\rightarrow & A^a_i + g_3\, a_i^a \qquad  A^a_i = 0 \nonumber\\
\Phi &\rightarrow & \Phi + g_3 \phi \nonumber\\
\Phi &=& \sqrt{\frac{1}{2}} {0 \choose \varphi} \qquad
\phi = \sqrt{\frac{1}{2}}
                {\chi^1 + i\chi^2 \choose  \eta  + i\chi^3} \quad.  \label{bgfl}
\end{eqnarray}
The fluctuations have to be gauge-fixed in some way. A class of covariant 
background gauges is given by
\be\label{gaugefix}
F^a = D_i(A)\, a_i^a + i \xi \left( 
    \Phi^\dagger \frac{\sigma^a}{2} \phi -
    \phi^\dagger \frac{\sigma^a}{2} \Phi 
    \right)
    = 0 \quad.
\ee
We use the 't~Hooft-Feynman gauge where the gauge parameter is $\xi = 1$. This 
is contrary to most publications on the electroweak phase transitions which 
work in the Landau gauge. 

Using eq.~(\ref{bgfl}) and eq.~(\ref{gaugefix}) the high temperature action 
(eq.~(\ref{Sht})) changes into
\be\label{Sbgfl}
S_{\rm ht} + \frac{1}{2 \xi} \int d^3x F^a F^a  \;\rightarrow\;
S_{\rm ht}^{\rm bg} + \delta S_{\rm lin} +
\delta S_{\rm quad} + \ldots
\ee
with the background field part
\be \label{Shtbg}
S_{\rm ht}^{\rm bg} = \frac{1}{g_3(T)^2}  \int d^3x \left[
\frac{1}{2} \partial_i\varphi \partial_i\varphi + 
 V_{\rm ht}(\varphi^2) \right] \quad.
\ee
The part which is linear in the fluctuations is proportional to $\eta$, the 
fluctuation which corresponds to the Higgs field $\varphi$ (cf.~eq.~(\ref{bgfl}))
\be
\delta S_{\rm lin} \propto  \frac{1}{g_3(T)} \eta \quad.
\ee
There are no linear terms in the other fluctuations. 
This is due to the fact that the background
field takes the minimum of this part of the fundamental action.
Note that we already made use of the absence of linear terms
in integrating out the non-static Matsubara-frequencies. Neglecting tadpoles here 
was only possible due to the fact that the remaining degrees of freedom are
purely 3-dimensional. 

The part quadratic in the fluctuations is most simply written in matrix 
notation 
\bea 
\delta S_{\rm quad} &=& \frac{1}{2} \int d^3x \; \; Q^T \cdot
\Bigg[ - \partial^2 {\bf 1} +
\left(
 \begin{array}{cccc}
   U & 0 & 0 & 0 \\
   0 & U & 0 & 0 \\
   0 & 0 & U & 0 \\
   0 & 0 & 0 & m_H^2
 \end{array}
\right) \Bigg]
\cdot Q \nonumber\\
Q^T &=& \left( a_1^1, a_2^1, a_3^1, \chi^1, \quad a_1^2, a_2^2, a_3^2, \chi^2,\quad
              a_1^3, a_2^3, a_3^3, \chi^3,\quad \eta \right) \quad. \label{Squad}
\eea
$U$ is a $4 \times 4$-matrix 
\bea
U &=&U_0 + \delta U \label{U}\\
U_0 &=& 
\left(
  \begin{array}{cccc}
     m_W^2 & 0 & 0 & 0 \\
     0 & m_W^2 & 0 & 0 \\
     0 & 0 & m_W^2 & 0 \\
     0 & 0 & 0  & m_\chi^2
   \end{array}
\right) 
\qquad
\delta U = 
\left(
  \begin{array}{cccc}
     0 & 0 & 0 & \partial_1\varphi \\
     0 & 0 & 0 & \partial_2\varphi \\
     0 & 0 & 0 & \partial_3\varphi \\
     \partial_1\varphi & \partial_2\varphi & \partial_3\varphi & 0
   \end{array}
\right) \quad. \label{U0}
\eea
The diagonal elements are the finite temperature squared masses
\bea
m_H^2 &=& \frac{\lambda_T}{g^2}\left( 3 \varphi^2 - 
        \left(\frac{v_0}{v}\right)^2 \right) \label{mH}\\
m_\chi^2 &=&  \frac{1}{4} \varphi^2 + 
            \frac{\lambda_T}{g^2}\left( \varphi^2 -
            \left(\frac{v_0}{v}\right)^2 \right) \\
m_W^2 &=&  \frac{1}{4} \varphi^2 \label{mW} \\
m_{\rm gh}^2 &=& \frac{1}{4} \varphi^2  \quad.
\eea
They are positive in the range of phase transition,
because $\left(\frac{v_0}{v}\right)^2$ is negative.
In the broken phase they are of the same order of magnitude
while gauge boson and ghost masses vanish in the symmetric
phase. Hence there is no mass hierarchy which holds 
over the whole interesting $\varphi$-range. 

Note that the $\eta$-fluctuation does not mix with the other 
fluctuations (eq.~(\ref{Squad})). In addition there are no linear terms
proportional to $a_i^a$ or $\chi^a$. In one-loop order
it is therefore possible to perform the remaining integrations in two steps. 
In a first step we integrate out gauge fields, ghosts,
and Goldstone bosons to get an effective action for the $\varphi$ field. We might 
not find a local effective action through this procedure, however,
because it is not supported by an appropriate  mass hierarchy. 
Nevertheless we shall discuss a local expansion, because the relevant 
scale is set by the bubble solutions which will be calculated in section 3.2.
Anticipating the results (figure 5) one sees that the corresponding mass
which is given by the inverse wall thickness is quite small 
($\approx \frac{1}{30}gv(T)$). 

In a second step, the fluctuations of the $\varphi$ field contributing to the static 
prefactor of the nucleation rate will be calculated.

\subsection{The effective action of the Higgs field $\varphi$}

The effective action of the Higgs field $\varphi$ is in one-loop order 
calculated from
\begin{equation}
S_{\rm eff}[\varphi] = S_{\rm ht}^{\rm bg}[\varphi] + \delta S[\varphi]
\end{equation}
with 
\begin{equation}\label{delS}
\delta S = \frac{1}{2}\log \det(-\partial^2 + M_{12}) - 
                     \log \det (-\partial^2 + M_{\rm gh}) \quad.
\end{equation}
The $12 \times 12$ $a$-$\chi$-matrix $M_{12}$ is a part of 
$\delta S_{\rm quad}$ (eq.~(\ref{Squad})), while the $3 \times 3$ ghost-matrix 
$M_{\rm gh}$ is easily calculated from the gauge-fixing condition
(eq.~(\ref{gaugefix})).
\be\label{ghmat}
M_{\rm gh} =
\left(
  \begin{array}{ccc}
     m_{\rm gh}^2 & 0 & 0 \\
     0 & m_{\rm gh}^2 & 0 \\
     0 & 0 & m_{\rm gh}^2
   \end{array}
\right)
\ee 
The derivative expansion can be carried out by calculating Feynman diagrams,
by summing up the relevant contributions of the heat-kernel expansion, or by 
using a method proposed in ref.~\cite{CarSal}. 
The latter is explained in more detail in appendix A.  The three methods 
of course yield identical results. 

One gets the effective action 
\be\label{Seff}
S_{\rm eff}[\varphi] =  \frac{1}{g_3(T)^2} \int d^3x \left[V_{\rm eff}(\varphi) 
     + \frac{1}{2} Z_H(\varphi) \partial_i\varphi \partial_i\varphi 
     + {\cal O}(\partial\varphi^4)\right]
\ee
with the effective potential
\bea
V_{\rm eff}(\varphi) &=& V_{\rm ht} - \frac{g_3(T)^2}{12 \pi} 
          \left( 9 m_W^3 - 6 m_{\rm gh}^3 + 3 m_\chi^3 \right) \label{Veff}\\
&=&  \frac{\lambda_T}{g^2} \left(\frac{1}{4}\varphi^4 - \frac{1}{2}
     \left(\frac{v_0}{v}\right)^2 \varphi^2 \right) \nonumber\\
&&  - \frac{g_3(T)^2}{12 \pi} \left( \frac{3}{8}\varphi^3 
    + 3 \left[ \frac{1}{4}\varphi^2 + \frac{\lambda_T}{g^2}\left( \varphi^2 -
     \left(\frac{v_0}{v}\right)^2 \right) \right]^{3/2} \right) \label{Veff2}
\eea
and the $Z$-function
\bea
Z_H(\varphi) &=&  1 \;+\;  \frac{g_3(T)^2}{4 \pi} \left[ 
   \;-\; \frac{3}{m_W + m_\chi}
   \;+\; \frac{3}{64} \frac{1}{m_W^3} \varphi^2 \right. \nonumber\\
 &&\qquad\qquad\qquad \left. -\; \frac{2}{64} \frac{1}{m_{\rm gh}^3} \varphi^2  
    \;+\; \frac{1}{64} \frac{1}{m_\chi^3} \left(1 + 
   4 \frac{\lambda_T}{g^2} \right)^2 \varphi^2 \right] \quad. \label{Zeff}
\eea 

The effective potential predicts a first-order phase transition. (We are 
restricting ourselves to zero-temperature Higgs masses $\tilde{m}_H \le
\tilde{m}_W$.) In the transition range it has two minima, the symmetric one at
$\varphi_S = 0$ and the asymmetric one at $\varphi_A = v(T)$. During the phase
transition $\varphi$ takes on values between $\varphi_S$ and $\varphi_A$. The 
most natural way of rescaling the field in eq.~(\ref{rescale}) 
is therefore to choose 
$v = v(T)$. Hence the asymmetric minimum is always at $\varphi_A = 1$. 

The parameter $\frac{v_0(T)^2}{v(T)^2}$ introduces the temperature dependence
into the potential.
The effective coupling $g_3(T)$ is related to this parameter by the 
requirement that the minimum of the potential corresponding to the 
broken symmetry phase is located at 
$\varphi_A = 1$, according to the field rescaling. 
The critical temperature $T_c$
corresponds to the two minima having equal height.  
At the roll-over temperature
$T_{\rm ro}$ the symmetric phase becomes unstable. In figure 1 we show  $\frac{v_0(T)^2}{v(T)^2}$ 
as function of $\frac{\lambda_T}{g^2}$, for $T=T_c$ and $T=T_{\rm ro}$, 
respectively. 
$g_3(T)^2$ at  $T=T_c$ and $T=T_{\rm ro}$ is shown in figure 2 .

\subsection{The gauge-fixing dependence of the effective action}

The effective potential (eq.~(\ref{Veff2})) is not a polynomial in $\varphi$,
due to the non-vanishing $\frac{\lambda_T}{g^2}$-term of the 
Goldstone mass (eq.~(\ref{Veff2})). Nevertheless, the corrections induced by 
this term are (at least for $\tilde{m}_H \le \tilde{m}_W$) numerically small. 
They manifest themselves in three (small) effects: \\
i) The critical temperature $T_c$ and the roll-over temperature $T_{\rm ro}$ 
are shifted towards higher temperatures. \\
ii) $V_{\rm eff}(\varphi=0)$ is shifted. This is corrected by adding a constant 
to the potential. We always work with $V_{\rm eff}(0) =0$. \\
iii) The effective Higgs mass defined by $\sqrt{V_{\rm eff}''(\varphi)}$ 
has different values in the symmetric and in the asymmetric phase. 
(cf.~our note on the Higgs masses in section 3.3)

Note that the contribution from the $\frac{\lambda_T}{g^2}$-term of the 
Goldstone mass is the only gauge-dependent part of $V_{\rm eff}(\varphi)$. 
Integrating out gauge bosons, Goldstones and ghosts in the limit of vanishing
$\frac{\lambda_T}{g^2}$ the effective potential is independent of the
gauge parameter $\xi$.  
The gauge-fixing dependent contributions due to 
the non-vanishing $\frac{\lambda_T}{g^2}$
are numerically small.

The situation is totally different looking at the $Z$-function (eq.~(\ref{Zeff})).
Comparing our results with those obtained in Landau gauge 
(see e.g.~\cite{BoBuFoHe}) the contribution from the mixed 
$W$-$\chi$ loops get a factor $\frac{2}{3}$ while
the sum of the $W$ and the ghost loops which has to be taken as one part 
gets a factor 10. The $\chi$-loop contribution is the same.

In figure 3 we have plotted the $Z_H$-function at the critical temperature 
in 't~Hooft-Feynman as well as in Landau gauge. The results are totally 
different. Note that this strong gauge-fixing dependence can not be cured by 
introducing a magnetic mass of any reasonable size. In the spirit of  ref.~\cite{Wei}
one might argue that the gauge-fixing dependence of $Z_H$ in one-loop
order cancels against the gauge-fixing dependence of $V_{\rm eff}$ in two-loop order. 
Indeed in the thin-wall limit, for $\frac{\lambda_T}{g^2}\rightarrow 0$,
one can see explicitly that higher derivative terms compete with lower 
derivative higher loop terms evaluated in the critical bubble background.
Away from this limits, however, we expect that also a cancellation 
against higher derivative terms of the same loop order plays an important role.

\subsection{Limits on derivative expansion and perturbation theory}

In calculating $S_{\rm eff}$ we have integrated out the gauge and Goldstone
degrees of freedom well separated from the Higgs field $\varphi$ in one-loop
order. In the range of small $\varphi$ these are massless or light modes
and one has to be aware of the breakdown of derivative expansion: the true
effective action is a non-local functional of $\varphi$. Indeed if one 
calculates higher derivative terms (beyond $(\partial \varphi)^2$) 
the singularities of the higher order $Z$-factors get worse.
If one inserts the critical bubble solutions obtained with the usual
kinetic term (and to be discussed in section 3.2) into the effective action
functional stated in derivative expansion
these terms diverge contrary to the $\int d^3x Z(\varphi) (\partial \varphi)^2$
term. The latter term, however, has defects as well. As mentioned above it is highly
gauge-dependent and $Z(\varphi)$ also turns out to become negative in some
$\varphi$-range in t'~Hooft-Feynman gauge. This makes it impossible to 
use the effective action resulting from the derivative expansion for a further 
treatment of the $\eta$ fluctuations. 

As an alternative to the derivative expansion, one may perform a heat kernel 
expansion. This is also a local expansion, but instead of summing contributions 
of a given number of derivatives to all orders, terms with different number of 
derivatives are systematically combined order by order. In this way, more and 
more non-locality is covered. There is no obvious scale and therefore
the dynamics of the symmetric phase is plagued by infrared problems. 
However, what we really want to know is a difference of the bubble 
effective action to the effective action of the symmetric phase.
The bubble solution provides additional scales which may serve as
infrared cut-off. As mentioned above (section 2.2) this scale is set 
by the inverse bubble wall thickness and turns out to be rather small.
Therefore, the heat kernel expansion may converge reasonably well
also in the presence of massless modes. This will not be attempted
in this paper. 

In order to take non-local effects into
account in some approximate way, we shall introduce some $\varphi$ independent
wave-function renormalization for the kinetic term of the Higgs field
\be\label{Sz}
S_z[\varphi(\vec x)]=\frac{1}{g_3(T)^2}\int d^3x \left[ \frac{1}{2} 
z(\partial_i\varphi)^2+V_{\rm eff}(\varphi(\vec x)) \right] \quad.
\ee
$z$ would have to be determined from the true non-local effective 
action in such a way that $S_z[\varphi]$
approximates that action locally in some neighborhood of the critical
bubble (but not globally in field space, which is not possible). 
Therefore, $z$ is not
directly related to the wave-function renormalization $Z_H(\varphi)$
(eq.~(\ref{Zeff})) but also summarizes the effect of all the higher derivative 
operators. Unfortunately, a fit of $z$ is technically involved and presently not feasible.
One should insert trial functions extremizing eq.~(\ref{Sz}) into the higher order
heat kernel expansion containing the $12 \times 12$ matrix potential. In the resulting terms the $z$-dependence factorizes out, and the total expression should be extremized with respect to $z$. 

One could also try to find some average $z_{\rm av}$ arguing that the small
$\varphi$-range is not important in the integrated action at least for small
$\tilde{m}_H$. In view of the negative $Z(\varphi)$-range we are sceptical 
about this procedure. Nevertheless we will present a plot of such a 
$z_{\rm av}$ in section 3.2. 

Note that the breakdown of the derivative expansion does not automatically imply
the breakdown of  perturbation theory. 
To inspect the latter question we calculated (Appendix A) the
$Z_{\rm gauge}(\varphi)$ prefactor of the gauge-kinetic term in one-loop
order. This term does not appear directly in the Higgs Lagrangian but
$g_3(T)^2/Z_{\rm gauge}(\varphi)$ is the effective gauge coupling at 
the scale $\varphi$. Figure 4  demonstrates that it becomes 
big ($Z$ small or negative, respectively) already halfway in $\varphi$
between the broken and unbroken minimum even for small $\tilde{m}_H$
($\tilde{m}_H=\frac{1}{2}\tilde{m}_W$).
This means that the perturbative expansion  breaks down in a rather big
range of $\varphi$ starting from $\varphi=0$, and that we cannot trust 
the perturbative potential and action for these Higgs field values
$\varphi$. This was already emphasized in ref.~\cite{Shapo}. It is not clear,
however, at what effective scale the gauge coupling will appear in higher 
loop order. Nevertheless, one has to expect a rather strong dependence 
of the bubble solutions on this part of the action (different from sphaleron
configurations  based on the broken phase \cite{HeKriSch}),
Also the potential in the ansatz eq.~(\ref{Sz}) might get 
nonperturbative contributions. They are not considered in this paper.

\section{The Phase Transition}
\setcounter{equation}{0}

\subsection{Nucleation rate}

According to the effective action calculated above the electroweak phase
transition is of first-order. It is triggered by bubble nucleation. Just
before the transition starts the system is in thermal equilibrium in the 
metastable symmetric phase. 

The onset of a first-order phase transition was investigated by Langer 
\cite{Langer} and successive work \cite{Affleck, Linde}
in some detail. The bubble nucleation at the electroweak phase transition 
has been investigated e.g.~in the references \cite{EnIgKaRu} and 
\cite{LiuMcLTu}.
It is assumed that the total system is dividable into interval
and system of interest \cite{Laser}. The influence of the heat bath on the 
rest is first to induce thermal fluctuations and second to change the free 
energy of the system of interest. Both influences are only taken into account 
on average. We divide the total system, which consists of all fields of the 
standard model, into the $\omega_0$-frequency
of the real scalar field $\varphi=\sqrt{2\Phi^\dagger\Phi}$
and the rest. The free energy is given by 
$\beta F[\varphi(\vec x)]=S[\varphi(\vec x)]$ where $S[\varphi(\vec x)]$
is the effective action. 

Langer solved the equations of motion in a 
neighborhood of the saddlepoint, which corresponds to the critical bubble, and 
got a quasi-stationary solution describing a density flow from the metastable 
to the stable region. The transition rate $\Gamma$ is the integral of density
flow and evaluates to
\be\label{rate}
\Gamma = 
\frac{\kappa}{2\pi}V\left(\frac{\bar{S}}{2\pi}\right)^{3/2}\frac{1}{\sqrt{
|\lambda_-|}}\left[\frac{\det''K}{\det K_0}\right]^{-1/2}\exp\{-\bar{S}\}
\ee
where

\begin{itemize}
\item $\kappa$ is the dynamical prefactor. It takes into account the dynamical
characteristics of the heat bath. We will not calculate it in this paper but refer to the 
literature \cite{CsKa}.
\item $V$ is the spatial volume.
\item
\be\label{Kdef}
K=g_3(T)^2\frac{\delta^2S}{\delta\varphi^2}\Bigr\vert_{\bar\varphi(\vec x)}
\qquad \qquad 
K_0=g_3(T)^2\frac{\delta^2S}{\delta\varphi^2}\Bigr\vert_{\varphi s}
\ee
$\bar\varphi(\vec x)$ is the critical bubble and $\varphi_s$ represents the
symmetric phase. $\det''K$ denotes the determinant of $K$, without the 
negative and the three zero eigenvalues. The negative eigenvalue corresponds
to the growing and shrinking of the critical bubble. The zero-modes are
due to translational invariance.
The factors $g_3(T)^2$ result from a 
rescaling which makes formulas more convenient. 
\item $\lambda_-$ is the negative eigenvalue of $K$.
\item $\bar{S}=S[\bar\varphi(\vec x)]-S[\varphi_s]$ is the effective action 
of the critical bubble.
\end{itemize}
$\bar{S},\lambda_-$ and the static prefactor
\be\label{prefac} 
A=\left[\frac{\det'' K}{\det K_0}\right]^{-1/2}
\ee
are functionals of the critical bubble $\bar\varphi(\vec x)$. The latter 
depends only on the temperature. The determinants 
now refer to the Higgs field fluctuations only. 

One aim of our work is to calculate the rate
\be\label{R} R=\frac{\Gamma}{V\kappa}=\frac{1}{2\pi}\left(\frac{\bar{S}}{2\pi}\right)^{3/2}
\frac{1}{\sqrt{|\lambda_-|}} \, A \, \exp\{-\bar{S}\}
\ee
as a function of the temperature.

We will describe our calculation for the zero-temperature Higgs mass 
$\tilde{m}_H=\frac{1}{2}\tilde{m}_W$  
in some detail and report the results for other values of $\tilde{m}_H$ 
at the end.

\subsection{The critical bubble}

The critical bubble $\varphi(\vec{x})$ is a saddlepoint of the effective action
$S[\varphi]$ and therefore a solution of
\be\label{ersteVar}
\frac{\delta S}{\delta\varphi}\Bigr\vert_{\bar\varphi(\vec x)}=0
\ee
with the boundary conditions
\be\label{3.6}
\lim_{\vec{x}\to\infty}\bar\varphi(\vec x)=\varphi_S
\qquad \bar\varphi(0)>\varphi_S \quad.
\ee
Taking the effective action $S_z[\varphi]$ the saddlepoint equation reads
\be\label{dgl}
z\partial^2\bar\varphi_z(\vec x)-V_{\rm eff}'(\bar\varphi_z(\vec x))=0 \quad.
\ee
By rescaling $\vec x\to\vec x/\sqrt z$ this equation reduces to the one
with $z=1$. The solution for arbitrary but constant $z>0$ is therefore
\be\label{phiz}
\bar\varphi_z(\vec x)=\bar\varphi\left(\frac{\vec x}{\sqrt z}\right)\qquad {\rm with}\qquad \bar\varphi=\bar\varphi_{z=1} \quad.
\ee
To cover the whole temperature range from the new roll-over temperature
$T_{\rm ro}$ to the critical temperature $T_c$ we 
introduce a temperature-like 
variable y instead  of $\frac{v_0(T)^2}{v(T)^2}$ via
\begin{eqnarray}\label{ydef}
&&\left(\frac{v_0(T)^2}{v(T)^2}\right)_y = 
\frac{v_0(T_c)^2}{v(T_c)^2} + y \left(\frac{v_0(T_{\rm ro})^2}{v(T_{\rm ro})^2} 
- \frac{v_0(T_c)^2}{v(T_c)^2} \right)
\nonumber\\[1.2ex]
&&y = 0.1, 0.2, \ldots, 0.9 \quad.
\end{eqnarray}
This way of dividing the interesting temperature interval turned
out to be much more appropriate than dividing it into equal $\Delta T$ intervals.
For a given value of $y$ the temperature $T$ and the effective action
(eq.~(\ref{Sz})) are determined; the temperatures are given in table 1.
We have calculated the critical bubbles from eq.~(\ref{dgl}) with $z=1$
for these nine $y$'s; they are spherical symmetric and plotted in figure 5.

The effective action of the 
critical bubble with arbitrary $z>0$ is, using eq.~(\ref{phiz})
\begin{eqnarray}
{\rm S_z}[\bar\varphi_z]&=&\frac{1}{g_3(T)^2}\int d^3x
\left[ \frac{1}{2} z \,(\partial_i\bar\varphi_z(\vec x))^2 + 
V_{\rm eff}(\bar\varphi_z(\vec x)) \right] \nonumber\\
&=&z^{3/2}\quad S_1[\bar\varphi(\vec x)]\label{Sscaling}
\end{eqnarray}
with
\be\label{S1}
S_1[\bar\varphi(\vec x)]=\frac{1}{g_3(T)^2}\int d^3x
\left[ \frac{1}{2} (\partial_i\bar\varphi(\vec x))^2 + 
V_{\rm eff}(\bar\varphi(\vec x))\right] \quad.
\ee
This simple scaling behavior clarifies the roll of the corrections to
the surface term and allows to proceed without knowing the value  of $z$ precisely.

Assuming that $V_{\rm eff}$ is renormalized by $V_{\rm eff}(0) =0$ we
identify $\bar{S}$ with  $S_1[\bar\varphi]$. It is given in table 1 and
plotted in figure 9 where single points are connected by a spline
(full line).

With a known critical bubble configuration one can evaluate 
\be 
z_{\rm av} = 
\frac{\int d^3x Z(\bar{\varphi}_z) (\partial \bar{\varphi}_z)^2}
     {\int d^3x (\partial \bar{\varphi}_z)^2} =
\frac{\int d^3x Z(\bar{\varphi}) (\partial \bar{\varphi})^2}
     {\int d^3x (\partial \bar{\varphi})^2}
\ee
in order to get an average $z_{\rm av}$. In figure 6 we have plotted it
for $\tilde{m}_H = \frac{1}{2} \tilde{m}_W$ versus $y$. However, in our 
opinion the range near $\varphi = 0$ is not taken into account properly
in this way, because $Z(\varphi)$ is unphysical, as argued in section 2.5.

\subsection{Eigenvalues of $K$ and $K_0$}

Using the effective action $S_z[\varphi]$ (eq.~(\ref{Sz})) the operators 
$K$ and $K_0$ defined in 
eq.~(\ref{Kdef}) are
\begin{eqnarray}
K &=& - z\partial^2 + U \qquad \qquad U = V_{\rm eff}''(\bar{\varphi}_z(r))
\label{K} \\
K_0 &=& - z\partial^2 + U_0 \qquad \quad U_0 = V_{\rm eff}''(\varphi_S) 
 = m_H(T)^2 \quad. \label{K0} 
\end{eqnarray}
$U_0$ is the squared effective mass of the Higgs field in the symmetric phase
and the natural mass-scale of the remaining effective action.

Note that there is a change in notation. $U$ and $U_0$ are in this chapter
no matrices as in eq.~(\ref{U0}) but real-valued functions of $\varphi$.
The Higgs mass in eq.~(\ref{mH}) corresponds to the high-temperature
action (eq.~(\ref{Sht})). It is $\varphi$ and, via $\frac{v_0}{v}$, 
temperature-dependent. $m_H(T)$ defined in eq.~(\ref{K0}) corresponds the 
effective action of eq.~(\ref{Sz}). It
is the effective Higgs mass in the symmetric phase (i.e.~at $\varphi_S$)
and hence only temperature dependent. In principle it is possible to 
define a $\varphi$-dependent effective Higgs mass via 
$m_H(T,\varphi)^2 = V_{\rm eff}''(\varphi)$, but this squared mass is
negative for some $\varphi$-values. We do not need it.

The negative eigenvalue of $K$ is determined by the eigenvalue equation
\be\label{nEuGl} 
K \,n(\vec x)=(-z\partial^2+V''(\bar\varphi_z(\vec x)) \,n(\vec x) 
= \lambda_- \,n(\vec x) \quad.
\ee
By rescaling $\vec x\to \sqrt z\vec x$ and using eq.~(\ref{phiz}) one sees that
$\lambda_-$ is independent
of $z$. Similarly all other eigenvalues of $K$ and $K_0$ are $z$-independent.
We calculated $\lambda_-$ numerically by solving the Schr\"odinger equation
(\ref{nEuGl}) with the boundary condition
\be\label{3.14}
\lim_{r\to\infty} n(r)=0 \quad.
\ee

The static prefactor $A$ (eq.~(\ref{prefac})) is a product of eigenvalues
of $K$ and $K_0$ and therefore independent of $z$. We evaluate it for
$z=1$.

\subsection{Heat-kernel method and calculation of the static prefactor}

Starting from 
\be\label{Schwinger}
\ln\left(\frac{\det K}{\det K_0}\right) =
-{\rm Tr}\int^\infty_0\frac{dt}{t}e^{-at}\left(
e^{-t(K-a)}-e^{-t(K_0-a)}\right)
\ee
it is possible to expand the logarithm in power of $t$ \cite{Carson}. 
A very elegant method to do this is provided by a new calculation scheme
\cite{FliSchmSchu}. One gets
\begin{eqnarray}\label{det2}
\ln\left(\frac{\det K}{\det K_0}\right)&=&-\sum^\infty_{n=1}\frac{1}{n!}
\int^\infty_0\frac{dt}{t}(4\pi t)^{-3/2}e^{-at}t^n(O_n(a)-O^{(0)}_n(a))
\nonumber\\
&=&-\sum^\infty_{n=1}\frac{\Gamma(n-3/2)}{n!}(4\pi)^{-3/2}a^{3/2-n}(O_n(a)-
O_n^{(0)}(a)) \quad.
\end{eqnarray}
The $O_n(a)$'s and $O^{(0)}_n(a)$'s are rather complicated functionals of 
$\bar\varphi(\vec x)$ which are given in Appendix C. They
depend on the mass scale $a$ pulled out in eq.~(\ref{Schwinger}). In doing the
$t$-integration for $n=1$ we have dimensionally regularized the UV-divergence
which may be traced back to the reduction in dimension \cite{CaMcL}. 
Note that pulling out the squared mass $a$ in eq.~(\ref{Schwinger}) 
regularizes the IR-divergencies and makes the $t$-integral finite at
the upper bound. This is an advantage over the method proposed in 
ref.~\cite{DPY}.

However eq.~(\ref{Schwinger}) is only valid for positive definite $K$ and $K_0$.
In our case, $K$ has one negative and three zero-eigenvalues. Exactly these
eigenvalues are left out of the static prefactor (eq.~(\ref{prefac})) of the
nucleation rate anyway. On the other hand, we have to drop four eigenvalues
of $K_0$ as well, if we want to use eq.~(\ref{Schwinger}), because this equation
makes use of the fact that the numbers of eigenvalues of $K$ and $K_0$ are 
equal.

Taking out four times the eigenvalue $U_0$ from $\det K_0$ one gets
\begin{eqnarray}
&& \hspace{-0.6cm} \ln\left( U_0^4 \, \frac{\det{}'' K}{\det K_0} \right) = 
\ln\left( \frac{\det{}'' K}{\det{}'' K_0} \right) \nonumber \\[0.5ex]
&=& \int_0^\infty \frac{dt}{t} \left[ - \left( {\rm Tr}\left\{e^{-t K} \right\}
    - e^{-t \lambda_-} - 3 \right) + 
      \left( {\rm Tr}\left\{ e^{-t K_0} \right\}
     - 4 e^{-t U_0} \right) \right] \nonumber \\[0.5ex]
&=& \int_0^\infty \frac{dt}{t} e^{-at}\left[ 
  - {\rm Tr}\left\{ e^{-t(K-a)} - e^{-t(K_0-a)} \right\} + \right. \nonumber \\ 
  && \qquad \qquad \qquad  \qquad
     \left. + \left( e^{-t(\lambda_- -a)} - e^{-t(U_0 -a)} \right) +
     3\left( e^{at} - e^{-t(U_0 -a)} \right) \right] \nonumber \\[0.5ex]
&=& \int_0^\infty \frac{dt}{t} e^{-at}\left[ 
    - \sum_{n=1}^\infty \frac{1}{n!} (4\pi)^{-3/2} t^{n-3/2}
       \left( O_n(a) - O_n^{(0)}(a) \right)+ \right. \label{drop0-}\\ 
  && \quad  \qquad
    + \sum_{n=1}^\infty \frac{1}{n!} t^n
       \left( (a-\lambda_-)^n - (a-U_0)^n \right) \left.
    + \, 3 \sum_{n=1}^\infty \frac{1}{n!} t^n
       \left( a^n - (a-U_0)^n \right) \right] \quad. \nonumber  
\end{eqnarray}
From the sum over the $O$'s only the first few terms are calculable. Therefore
we have to truncate the other two sums as well. From similar calculations with
other models we found  that the truncation is best done at the 
`same' powers of $t$ rather than at the same number $N$ of terms
(see also \cite{Schechter, Fliegner}). This cannot be done straightforwardly
however, because the $O$-sum runs over half-integer 
power of $t$ while the other sums run over full-integer powers. We solved the 
problem by defining the two functions
\begin{eqnarray} 
X(N,a) &=& \int^\infty_0 \frac{dt}{t} e^{-at} \left[-\sum_{n=1}^N \frac{1}{n!}
(4 \pi)^{-3/2} t^{n-3/2} \left( O_n(a) - O_n^{(0)}(a) \right)\right] 
\nonumber \\
&=& - \sum_{n=1}^N \frac{\Gamma (n-3/2)}{n!} (4\pi)^{-3/2}a^{3/2-n}
\left( O_n(a) - O_n^{(0)}(a) \right) \label{X} 
\end{eqnarray}
and 
\begin{eqnarray} 
Y(N',a) &=& \int_0^\infty \frac{dt}{t} e^{-at} \sum_{n=1}^{N'} \frac{1}{n!}
t^n \left((a-\lambda_-)^n + 3a^n - 4(a-U_0)^n \right) \nonumber \\
&=& \sum_{n=1}^{N'} \frac{1}{n} \left[ \left(\frac{a-\lambda_-}{a} \right) ^n
+ 3 -4\left(\frac{a-U_0}{a} \right) ^n \right] \quad. \label{Y}
\end{eqnarray}
While we have been able to evaluate $X(N,a)$ for $N \in \{1,2,3,4,5,6\}$ 
\footnote{One could evaluate $X(7,a)$ as well \cite{FliSchmSchu}, 
but this does not appear necessary in view of the excellent convergence.}, 
$Y(N',a)$ could be calculated
for every integer $N'$. We interpolated $Y(N',a)$ by a spline and defined
the functions
\begin{equation}\label{WNdef}
W_N(a) = X(N,a) + Y(N-\frac{3}{2}, a) \quad.
\end{equation}
From the equations (\ref{prefac}, \ref{drop0-} -- \ref{WNdef}) 
\begin{equation} \label{LimWN}
\lim_{N \rightarrow \infty} W_N(a) = \ln \left(U_0^4 \frac{\det''K}{\det K_0}
\right) = - 2 \ln \left(\frac{A}{m_H(T)^4} \right)
\end{equation}
follows. While the $W_N(a)$'s are functions of $a$, the limit is not. 
This will give us a good criterion for the quality of convergence \cite{Laser}.

For every critical bubble calculated above we have evaluated the functions $W_1(a)$, \ldots, $W_6(a)$. Figure 7 shows the typical behavior.
If $a$ -- the squared mass pulled out -- is too small, there is no convergence
at all. If $a$ is too big, the convergence is bad. But if $a$ is similar
to the natural mass scale $U_0$, the functions converge quite well towards
a constant, which is plotted as dashed line.

There are several sources of error in the outlined procedure:
numerical errors, ambiguities in interpolating $Y(N',a)$ and 
uncertainties in fixing the limit of the $W_N(a)$'s. We have estimated 
them to be less than 2\% . The values of $\ln(A/T^4)$ 
are listed in table 1 and plotted in figure 9 (dashed line).
 
\subsection{The static prefactor and the effective 
             action of the critical bubble}

Up to now the static prefactor
has usually been estimated from dimensional reasons as $T^4$ \cite{Linde},
or as $m_H(T)^4$ \cite{LiuMcLTu}. There are some other calculations of the
prefactor, which take only into account the lowest eigenvalues of $K$ 
\cite{BuFoHeWa, CoKaMau}.
The results of these calculations are somewhere between the two dimensional
estimates. To compare our results with these values we have expressed $A$
in units of GeV.
In figure 8 they are plotted together 
with these two estimates. One sees that the static prefactor calculated 
by us varies substantially from $y=0.1$ to $y=0.9$. On the side of the 
critical temperature $(y=0)$ it is much smaller than previous values.
This results in a  much smaller nucleation rate.

In a recent paper \cite{BaKi} the fluctuation corrections to 
critical bubbles calculated from the usual model effective potential are 
discussed. The method is based on the solution of Schr\"odinger type 
eigenvalue equations and totally different from our procedure. Unfortunately 
the present results are hard to compare because in ref.~\cite{BaKi} the 
high-temperature limit is not taken and because the potential differs in 
detail. A comparison of the two methods in a common case would be very 
interesting.

The logarithm of the rate $R/T^3$ defined in eq.~(\ref{R}) is
\begin{equation}
\ln \left( \frac{R}{T^3} \right) = 
    \ln \left( \frac{1}{2\pi}\left(\frac{\bar{S}}{2\pi}\right)^{3/2}
                \frac{T}{\sqrt{|\lambda_-|}} \right)
  + \ln \left( \frac{A}{T^4} \right) - \bar{S} \quad.
\end{equation}
The different contributions are plotted in comparison in figure 9. One sees that 
the first one is small and
nearly constant. The nucleation rate is determined by $-\bar{S}$ and 
$\ln(A/T^4)$. The comparison of these two values
decides on the reliability of our results. $-\ln(A/T^4)$ is nothing but 
the one-loop correction to $\bar{S}$ coming from scalar loops. Therefore 
$|\ln(A/T^4)|\ll \bar{S}$ is required for consistency. 

Very close to the critical and the roll-over temperature 
it matters that the radiative corrections
shift the values of $T_c$ and $T_{\rm ro}$. This effect should perhaps better
be incorporated in the quasiclassical effective action. In our approach it
causes an increase of the static prefactor near these temperatures. However,
investigating the electroweak phase transition the interesting temperatures 
are well separated from $T_c$ and $T_{\rm ro}$, as we will see below. 

In Langer's theory the static prefactor $A$ takes the possibility
into account that the phase transition may be started by a bubble which
differs from the critical one. On the other hand the whole theory is based on
a solution of the equations of motion in the neighborhood of the 
saddlepoint that corresponds to the critical bubble. From this reasoning
one again gets that $|\ln(A/T^4)|\ll \bar{S}$ should be valid.
This comparison is not unambiguous because $\bar{S}$ is dimensionless while 
$A$ is not. Expressing $A$ in $T$ seems to be appropriate because the
temperature is the typical scale of the phase transition.

Therefore the nucleation rate calculated by us is not reliable at temperatures
near the roll-over temperature $y=1$. Taking estimates based on
cosmological reasons however, the electroweak phase transition starts when 
$\ln(R/T^3)\approx 140$. \footnote{Here we have assumed
that $\kappa={\cal O}(1)$.} The relative starting temperature is therefore $y_s=0.42$.
At this temperature $|\ln(A/T^4)| \approx 0.3 \bar{S}$. This
is an acceptable correction. 

However, with $z$ smaller than 1 we find according to eq.~(\ref{Sscaling})
an additional suppression of the 
leading term ($\bar{S}$), i.e. the the one-loop contribution 
becomes relatively more important, indicating a less convergent loop expansion.

\subsection{Results for other Higgs masses}

The numerical calculations presented have been done for the zero 
temperature Higgs masses $\tilde{m}_H=\frac{1}{2}\tilde{m}_W, 
\frac{3}{4}\tilde{m}_W$ and $\tilde{m}_W$. The results are listed in 
the tables 1, 2 and 3. 

The critical temperature $T_c$ and the roll-over temperature $T_{\rm ro}$
depend on the Higgs mass. One should compare values corresponding to different 
masses at the same relative temperature $y$ defined in eq.~(\ref{ydef}).
The actions $\bar{S}$ of the critical bubbles depend strongly on the Higgs 
mass, while the Higgs mass dependence of the static prefactor is only
small. Although the rates are quite different for the three masses
the relative onset temperatures differ only moderately. This is
due to the fact that the rates decrease rapidly when the temperature is
lowered. 

We have listed the relative onset temperatures in table 4 
together with the corresponding values of
$\bar{S}$, $\ln(A/T^4)$ and the quotient of both.
This quotient decides, as argued above, on the reliability of the
nucleation rate formula (eq.~(\ref{rate})). 
One sees that the corrections get worse if the Higgs mass increases.
 
\subsection{Comparison with the thin-wall approximation}

If the temperature is just below the critical temperature $T_c$
the radius $R$ of the critical bubble is much larger than the size of
the bubble wall $\Delta R$. In the limit $T \rightarrow T_c$ and
$\Delta R \ll R$ the effective action of the critical bubble can be 
written as (for $z=1$)
\begin{equation}
\bar{S}_{\rm TW} = \frac{1}{g_3^2} \frac{16 \pi \sigma^3}{3 \epsilon^2}
\end{equation}
with the volume energy
\be
\epsilon = - V_{\rm eff}(T,\varphi_A)
\ee
and the surface tension
\be
\sigma = \int^{\varphi_A}_{\varphi_S} d\varphi 
         \sqrt{2V_{\rm eff}(T_c,\varphi(r))}
\ee
(cf.~appendix B). 
The values of $\sigma$ are given in table 5. The values of $\bar{S}_{\rm TW}$
are listed in comparison with the full numerical values of the critical
bubbles effective actions. (See figure 10 as well.) The shape 
of the critical bubbles is far from thin-wall except in the case $y=0.1$
(cf.~figure 5, for other Higgs masses the bubbles are similar). 
Nevertheless the thin-wall estimates of the corresponding effective actions 
are quite good even for smaller temperatures. 
On the other hand one has to calculate the critical
bubble configurations themselves if one wants to evaluate the static 
prefactor.

\section{Discussion and Conclusions}

Our main result is a very accurate determination of the Higgs fluctuation determinant
for critical bubble solutions for an action (eq.~(\ref{Sz})). It leads to a 
rather drastic change in the prefactor $A$ of the transition rate compared to 
previous rough estimates as indicated in figure 8. The transition is 
suppressed stronger. It is interesting to note
that the radiative corrections do not depend on the normalization factor $z$
while the quasiclassical bubble action scales with $z^{3/2}$.
In our procedure to evaluate the heat kernel expansion it was essential that we 
go to a rather high order, that we 
take out a variable scale and that we treat the subtraction of the unstable
and zero modes carefully. The separate interpolation for integer and half integer
$O_n$ before the subtraction is essential for the quality of the approximation.

We have also discussed  the critical bubble configurations and their action
for various temperatures between $T_c$ and $T_{\rm ro}$, and
Higgs masses. For $\tilde{m}_H \leq \frac{3}{4}\tilde{m}_W$ and temperatures 
not too close to $T_{\rm ro}$ the radiative corrections are small compared 
to the $z=1$ quasiclassical action
term (see table 2). Surprisingly the thin-wall approximation for the action
is quite good at temperatures where the critical bubble profile is not
`thin-wall' any more. The dependence of the action on the constant $Z$-factor 
can be taken out explicitly. It is an interesting side remark that the usual 
thin-wall machinery goes through even for arbitrary positive factors
$Z(\varphi)$ (Appendix B).

In the first part of the paper we discussed the status of a perturbative
effective Higgs action. The 't~Hooft-Feynman covariant background gauge
is particularly well suited for a discussion of the separate integration
of gauge, Goldstone, and ghost fields, and also avoids IR problems near
the broken phase vacuum. The size of the rescaled 3-dimensional gauge
coupling $g_3(T)^2$ in the broken phase changes smoothly in the phase
transition and is not very big and not very small; thus both the perturbative
expansion in $g^2_3$ and the high temperature expansion seem to work. The
inspection of the one-loop $Z$ prefactors of the Higgs and gauge-kinetic terms
tells us however, that this is only true in the broken phase. 

Even at small $\tilde{m}_H\sim\frac{1}{2}\tilde{m}_W$
the Higgs $Z_H(\varphi)$ (figure 3) becomes negative already at rather big 
values of $\varphi$ of the Higgs fields in our gauge. Different from the 
one-loop potential, $Z_H(\varphi)$ is very gauge-dependent and differs from 
the Landau gauge. The negativity of $Z_H$ cannot be changed by the 
introduction of a reasonable magnetic mass. The latter
can only change the singular behavior of $Z_H(\varphi)$ for $\varphi$
very close to zero. As we argued in chapter 3, this signals the breakdown
of the derivative expansion. 

Even worse, the gauge field $Z_{\rm gauge}(\varphi)$ becomes negative at still 
larger values of $\varphi$. Since it gives the static $\varphi$-dependent 
effective gauge coupling\footnote{We have checked that this behavior gives
the bound $\frac{m_W(T,\varphi)}{T}>0.1$ for a perturbative treatment
to be compared with $\frac{m_W(T,\varphi)}{T}>0.2$ from a renormalization
group calculation \cite{ReuWe2}.} $g_3(T)^2/Z_{\rm gauge}(\varphi)$, this 
signals the breakdown of perturbation theory for a big range of $\varphi$
already at small $\tilde{m}_H\sim\frac{1}{2}\tilde{m}_W$
(figure 4), though the optimal choice of a scale of a gauge coupling
in multi-loop calculations in the background of a critical bubble is not 
known. Thus,  the use of a Higgs action (eq.~(\ref{Sz})) inspired by 
perturbation theory in the discussion of the
critical bubbles does not have a firm ground. The `small' $\varphi$ 
region contributes directly to the critical bubble action through the 
Higgs kinetic term and not so much through the potential, but of course 
there is an indirect effect of the potential changing (perhaps drastically) 
the critical bubble configuration itself.
 
\section*{Acknowledgments}
We would like to thank W.~Buchm\"uller and C.~Wetterich for useful discussions.

\newpage

\section*{Appendix A: 
Calculation of the Wave Function Renormalization $Z$ Factors} 
\setcounter{equation}{0}
\renewcommand{\theequation}{A.\arabic{equation}} 

In this Appendix we calculate the $Z$ factors in front of the kinetic
terms. We work in 't~Hooft-Feynman background gauge (eq.~(\ref{gaugefix})).

We first 
assume that there is only a real scalar background field (eq.~(\ref{bgfl})). 
The task is to expand $\delta S$ (eq.~(\ref{delS})) in powers of $\partial \varphi$.
It reduces to the calculation of
\be\label{Wld}
W = \log \det (-\partial^2 + U) = \frac{1}{3} \log \det (-\partial^2 + M_{12})
\ee
where $U$ is given in eq.~(\ref{U}). Following \cite{CarSal} we expand
\be\label{W}
W = \int \frac{d^3k}{(2 \pi)^3} \left[ \int d^3 x {\rm tr} \log (\Delta^{-1}) + 
\frac{k^2}{3} \int d^3x {\rm tr} (\partial_i \Delta)^2 + \ldots \; \right]
\ee
where
\be
\Delta = (k^2 + U(\varphi(\vec{x}))^{-1} \quad.
\ee
tr denotes the trace over the $4 \times 4$-matrices. The logarithm
is expanded as
\begin{eqnarray}
{\rm tr} \log (\Delta ^{-1}) &=& {\rm tr} \log (k^2 + U_0 + \delta U) 
      \nonumber \\
&=& {\rm tr} \log \left(\Delta_0^{-1} (1 + \Delta_0 \delta U)\right) 
      \nonumber \\
&=& {\rm tr} \left[\log(\Delta_0^{-1}) + \sum_{n=1}^{\infty} 
    \frac{(-1)^n}{n} (\Delta_0 \delta U)^n \right]
\end{eqnarray}
where $U_0$ and $\delta U$ are given in eq.~(\ref{U0}) and
\be
\Delta_0 = (k^2 + U_0)^{-1} = 
\left( \begin{array}{cccc}
(k^2 + m_W^2)^{-1} & 0 & 0 & 0 \\
0 & (k^2 + m_W^2)^{-1} & 0 & 0 \\
0 & 0 & (k^2 + m_W^2)^{-1} & 0 \\
0 & 0 & 0 & (k^2 + m_\chi^2)^{-1} \\
\end{array} \right) 
\ee
Calculating the $Z$-function only the terms proportional to 
$(\partial \varphi)^2$ are of interest. Hence only the $n = 2$ term contributes, and
the relevant part of the first term on the r.h.s.~of eq.~(\ref{W}) is 
\begin{eqnarray}
&& \int \frac{d^3k}{(2 \pi)^3} \int d^3x \frac{1}{2} 
   {\rm tr} (\Delta_0 \delta U \Delta_0 \delta U)  \nonumber\\
&=& \int \frac{d^3 k}{(2 \pi)^3} \int d^3 x \frac{1}{k^2 + m_W^2}
    \frac{1}{k^2 + m_\chi^2} (\partial_i \varphi)^2 \nonumber\\
&=& \int d^3x \frac{1}{4 \pi} \frac{1}{m_W + m_\chi} (\partial_i \varphi)^2 \quad.
     \label{1stterm}
\end{eqnarray}
To evaluate the second term on the r.h.s.~of eq.~(\ref{W}) we make use of the
fact that
\be
\Delta = \Delta_0 + {\cal O}(\partial \varphi) \quad.
\ee
Since we are only interested in terms proportional to $(\partial \varphi)^2$
we may replace $\Delta$ by $\Delta_0$. The latter is diagonal and the trace 
tr reduces to a sum over the field degrees of freedom. For one degree one gets
\begin{eqnarray}
&&\int \frac{d^3k}{(2\pi)^3} \frac{k^2}{3} \int d^3x (\partial_i 
(k^2 + m^2)^{-1})^2 \nonumber \\
&=& \int d^3x \frac{1}{3} \int \frac{dk}{(2 \pi)^3} \left( \frac{-k^2}{(k^2 + m^2)^2}
\left( \frac{\partial m^2}{\partial \varphi} \right) \partial_i \varphi 
\right)^2 \nonumber \\
&=& \int d^3x \frac{1}{192\pi} \frac{1}{m^3} 
    \left( \frac{\partial m^2}{\partial \varphi} \right)^2
(\partial_i \varphi)^2 \quad. \label{2ndterm}
\end{eqnarray}
The ghost calculation is exactly the same 
(cf.~eq.~(\ref{delS}) and eq.~(\ref{ghmat}) ).

Finally one has to put the parts together and get the factors right.
Eq.~(\ref{2ndterm}) has to be summed over 9 gauge fields components, 
3 Goldstones and 3 ghosts.
Eq.~(\ref{1stterm}) has to be multiplied by 3 from eq.~(\ref{Wld}). 
From eq.~(\ref{delS}) one gets a factor $\frac{1}{2}$  respectively $-1$. 
The result is
\bea
&&\int d^3x \; \frac{1}{2}\; \left[
   -\; \frac{3}{4 \pi} \frac{1}{m_W + m_\chi}
    \;+\; \frac{3}{64 \pi} \frac{1}{m_W^3} 
   \left(\frac{\partial m_W^2}{\partial \varphi} \right)^2\right. \nonumber\\
&&\qquad\qquad \left. \quad-\; \frac{2}{64 \pi} \frac{1}{m_{\rm gh}^3} 
      \left(\frac{\partial m_{\rm gh}^2}{\partial \varphi} \right)^2 
    \;+\; \frac{1}{64 \pi} \frac{1}{m_\chi^3} 
      \left(\frac{\partial m_\chi^2}{\partial \varphi} \right)^2 
\right] (\partial_i \varphi)^2 \quad.
\eea
Taking into account the factor $\frac{1}{2g_3^2}$ of eq.~(\ref{Seff}) one gets
$Z_H(\varphi)$ (eq.~(\ref{Zeff})). 

Integrating out the Higgs field as well, the effective potential 
(eq.~(\ref{Veff})) gets an additional $-g_3(T)^2\frac{1}{12 \pi} m_H^3$ 
term while the $Z_H$-factor (eq.~(\ref{Zeff})) is modified by
\be
+\; \frac{1}{192 \pi} g_3(T)^2 \frac{1}{m_H^3} 
      \left(\frac{\partial m_H^2}{\partial \varphi} \right)^2 \quad.
\ee

With the complex scalar doublet and the gauge field as background fields
the same method gives a $Z$-function in front of the kinetic Goldstone term 
\be\label{A.7}
Z_\chi = 1 - g_3(T)^2 \frac{1}{24\pi} 
 \left(\frac{\lambda}{g^2} - \frac{1}{8}\right)^2 \varphi^2m_\chi^{-3} - 
g_3(T)^2  \frac{1}{4\pi} \left( \frac{2}{m_W + m_\chi} + \frac{1}{m_W + m_H} \right)
\ee
and in front of $\frac{1}{4} F_{ij}^a F_{ij}^a$
\be 
Z_{\rm gauge} = 1 - g_3(T)^2 \frac{1}{8 \pi} \left[ \frac{7}{m_W} - 
         \frac{1}{8m_\chi} - \frac{1}{24m_H} + \frac{1}{3m_{\rm gh}} \right] \quad.
\ee
The logarithmic derivative of $Z_{\rm gauge}$ is the $\beta$-function of the 
theory. $Z_{\rm gauge}$ is plotted in figure 4 for 
 $\tilde{m}_H = \frac{1}{2} \tilde{m}_W$,
$\frac{3}{4} \tilde{m}_W$ and $\tilde{m}_W$ versus $\varphi$. 

\section*{Appendix B: The Thin-Wall Approximation}
\setcounter{equation}{0}
\renewcommand{\theequation}{B.\arabic{equation}} 

If the extension $\Delta R$ of the bubble wall is small compared to
the radius $R$ of the critical bubble its effective action 
$S_1[\bar{\varphi}(r)]$ (eq.~(\ref{S1})) may be written as
\begin{eqnarray}
S_{\rm TW}[\bar{\varphi}(r)] &=&
\frac{1}{g_3(T)^2} 4\pi \left[
\int\limits_0^{R-\Delta R/2} dr r^2 V(\bar{\varphi}(r))
+ \int\limits_{R-\Delta R/2}^{R+\Delta R/2} dr r^2 
      \left[ \frac{1}{2}(\partial_r \bar{\varphi}(r))^2 + V(\bar{\varphi}(r))
      \right] \right] \nonumber\\
&\approx& \frac{1}{g_3(T)^2} \left[ 4\pi R^2 \sigma - \frac{4\pi}{3} R^3
\epsilon \right] \label{STWinitial}
\end{eqnarray}
with
\begin{eqnarray}
\epsilon &=& -V(\varphi_A)\\
\sigma &=& \int_{R-\Delta R/2}^{R+\Delta R/2} dr 
      \left[ \frac{1}{2}(\partial_r \bar{\varphi}(r))^2 + V(\bar{\varphi}(r))
      \right] \quad.  
\end{eqnarray}
In this appendix $V(\varphi)$ designates the effective potential.
We assume that it is normalized by $V(0)=0$.

Using $\Delta R \ll R$ again the saddlepoint equation (\ref{dgl}) 
reads ($z$=1)
\be
\frac{d^2 \bar{\varphi}(r)}{dr^2} = V'(\bar{\varphi}(r)) \quad.
\ee
Solutions of this equation have the `constant of motion'
\be\label{SolutionTW}
\frac{1}{2}\left(\frac{d \bar{\varphi}(r)}{dr}\right)^2 - V(\bar{\varphi}(r))
\ee
which is equal to $V(r\rightarrow \infty) = V(\varphi_S) = 0$ due to 
boundary conditions. Hence the surface tension may be written as
\be\label{sigma}
\sigma = \int^{\varphi_A}_{\varphi_S} d\varphi \sqrt{2V(\varphi(r))} \quad.
\ee
This integral is only real in the limit $T\rightarrow T_c$ where the 
derivation is exact.

Maximizing the thin-wall effective action with respect to the radius one gets 
the critical radius
\be
R_c = \frac{2 \sigma}{\epsilon}
\ee
and the effective action of the critical bubble as function of the 
surface tension $\sigma$ and the volume energy $\epsilon$
\begin{equation}\label{STW}
\bar{S}_{\rm TW} = \frac{1}{g_3^2} \frac{16 \pi \sigma^3}{3 \epsilon^2}
\end{equation}
which has been evaluated in section 3.7 .

Using that $V(\varphi)$ scales with $\frac{\lambda_T}{g^2}$ in leading
order one gets
\be
\epsilon \propto  \frac{\lambda_T}{g^2} \qquad
\sigma \propto \left( \frac{\lambda_T}{g^2} \right)^{1/2} \qquad
\bar{S}_{\rm TW} \propto \left( \frac{\lambda_T}{g^2} \right)^{-1/2} \quad.
\ee

The thin-wall approximation  may even be done analytically for the
effective action 
\be\label{SZphi}
S[\varphi(\vec x)]=\frac{1}{g_3(T)^2}\int d^3x
 \left[\frac{1}{2}Z(\varphi) (\partial _i\varphi)^2+V(\varphi(\vec x))
 \right] 
\ee 
with a general $\varphi$-dependent positive $Z(\varphi)$. 
The saddlepoint equation reads in this case
\be
Z(\bar{\varphi})\partial^2\bar{\varphi}+\frac{Z'(\bar{\varphi})}{2}
(\partial \bar{\varphi})^2 = V'(\bar{\varphi}) \quad.
\ee
With the substitution 
\be
V(\varphi)=\tilde V(\varphi)\cdot Z(\varphi)
\ee
we obtain
\be
Z(\bar{\varphi})\partial^2\bar{\varphi} = 
Z(\bar{\varphi}) \tilde{V}'(\bar{\varphi}) 
+ Z'(\tilde{V}'-\frac{1}{2}(\partial\bar{\varphi})^2)
\ee
which is in thin-wall approximation  solved by
\be
\frac{d\bar{\varphi}}{dr}=\sqrt{2\tilde{V}(\bar{\varphi}(r))} \quad.
\ee
Repeating the steps from eq.~(\ref{STWinitial}) to eq.~(\ref{STW}) but using this 
solution instead of eq.~(\ref{SolutionTW}) we one may bring eq.~(\ref{SZphi})
into the form eq.~(\ref{STW}), but with 
\be
\sigma_Z = \int^{\varphi_A}_{\varphi_S} d\varphi 
            \sqrt{2 Z(\varphi) V(\varphi(r))}
\ee
instead of $\sigma$ of eq.~(\ref{sigma}). For constant 
$Z(\varphi) = z$ this changes $S_{\rm TW}$ by a factor $z^{3/2}$
as it should according to eq.~(\ref{Sscaling}).

\section*{Appendix C: Operators of the Heat Kernel Expansion}
\setcounter{equation}{0}
\renewcommand{\theequation}{C.\arabic{equation}} 

The first six of the functionals $O_n(a)$ respectively $O_n^{(0)}(a)$
introduced in eq.~(\ref{det2}) are \cite{CaMcL,FliSchmSchu}
\bea
O_1 &=& \int dx \biggl(  U \biggr) \cr
\noalign{\vskip8pt}
O_2 &=& {1\over 2!} \int dx \biggl( U^2 \biggr) \cr
\noalign{\vskip8pt}
O_3 &=& {1\over 3!} \int dx
\biggl(  U^3
        + {1\over 2} \deli U \deli U
\biggr) \cr
\noalign{\vskip8pt}
O_4 &=& {1\over 4!} \int dx
\biggl( U^4
        + 2 U \deli U \deli U
        + {1\over5} \delij U \delij U \biggr) \cr
\noalign{\vskip8pt}
O_5 &=& {1\over 5!} \int dx \biggl(
        U^5
       + 3 U^2 \deli U \deli U
       + 2 U \deli U U \deli U
       + U \delij U \delij U
       + {5\over 3} \deli U\delj  U \delij U \cr
&&\qquad\qquad       + {1\over 14} \delijk U \delijk U
\biggr)\cr
\noalign{\vskip8pt}
O_6 &=& {1\over6!} \int dx
  \biggl( U^6
      + 4 U^3 \deli U \deli U
      + 6 U^2 \deli U U \deli U
      + {12\over7} U^2 \delij U \delij U\cr
&&\qquad\qquad      + {9\over7} U \delij U U \delij U
     + {26\over7} U \delij U \deli U \delj U
  + {26\over7} U \deli U \delj U \delij U\cr
&&\qquad\qquad      + {17\over14} \deli U \delj U \deli U \delj U
     + {18\over7} U \deli U \delij U \delj U
      + {9\over7} \deli U \deli U \delj U \delj U\cr
&&\qquad\qquad   + {3\over7} U \delijk U \delijk U
      + \delk U \delij U \delijk U
      + \delk U \delijk U \delij U\cr
&&\qquad\qquad      + {11\over21} \delij U \deljk U \delki U
      +  {1\over42} \delijkl U \delijkl U \biggr)
\eea
where $U$ is to replace by $U-a$ respectively $U_0 -a$. 
$U$ and $U_0$ are given in eq.~(\ref{K}) and eq.~(\ref{K0}).

\newpage


\newpage

\section*{Tables}

\begin{table}[ht]
\centering
\begin{tabular}{|r|r|r|r|r|r|}
\hline
$y$ & $T$ & $\frac{\lambda_-}{(g v(T))^2}$ & $\bar{S}$ & 
    $\ln\left(\frac{A}{T^4}\right)$ & $\ln\left(\frac{R}{T^3}\right)$ \\[1.2ex]
\hline
0.1 & 72.58 & -0.09 $10^{-3}$ & 2216.4 & -154.~ & -2358.~\\  
0.2 & 72.56 & -0.37 $10^{-3}$ & 558.1  & -83.9 & -632.6\\ 
0.3 & 72.53 & -0.84 $10^{-3}$ & 244.3  & -50.0 & -286.4\\ 
0.4 & 72.49 & -1.55 $10^{-3}$ & 131.5  & -35.6 & -160.5\\ 
0.5 & 72.44 & -2.53 $10^{-3}$ & 77.3   & -28.4 & -100.2\\ 
0.6 & 72.38 & -3.68 $10^{-3}$ & 46.4   & -24.3 & -66.2\\ 
0.7 & 72.30 & -4.63 $10^{-3}$ & 26.8   & -21.9 & -45.2\\ 
0.8 & 72.21 & -4.71 $10^{-3}$ & 13.6   & -20.7 & -31.8\\ 
0.9 & 72.11 & -3.23 $10^{-3}$ & 4.8    & -20.7 & -24.5\\
\hline
\end{tabular}
\caption{Numerical results for $\tilde{m}_H = \frac{1}{2}\tilde{m}_W$}
\end{table}

\begin{table}[h]
\centering
\begin{tabular}{|r|r|r|r|r|r|r|r|r|r|}
\hline
$y$ & $T$ & $\frac{\lambda_-}{(g v(T))^2}$ & $\bar{S}$ & 
    $\ln\left(\frac{A}{T^4}\right)$ & $\ln\left(\frac{R}{T^3}\right)$ \\[1.2ex]
\hline
0.1 & 97.27 & -0.14 $10^{-3}$  & 1086.9 & -115.~ & -1191.~\\ 
0.2 & 97.25 & -0.58 $10^{-3}$  & 273.8 & -82.0 & -347.0\\ 
0.3 & 97.23 & -1.33 $10^{-3}$  & 120.0 & -50.6 & -163.5\\ 
0.4 & 97.19 & -2.45 $10^{-3}$  & 64.8 & -36.7 & -95.6\\ 
0.5 & 97.15 & -4.00 $10^{-3}$  & 38.2 & -29.5 & -63.0\\ 
0.6 & 97.10 & -5.86 $10^{-3}$  & 23.1 & -25.4 & -44.8\\ 
0.7 & 97.04 & -7.44 $10^{-3}$  & 13.5 & -23.0 & -33.8\\ 
0.8 & 96.97 & -7.65 $10^{-3}$  & 7.0 & -21.8 & -27.1\\ 
0.9 & 96.87 & -5.21 $10^{-3}$  & 2.6 & -22.0 & -24.2\\
\hline
\end{tabular} \label{tab07}
\caption{Numerical results for $\tilde{m}_H = \frac{3}{4}\tilde{m}_W$}
\end{table}

\begin{table}[hb]
\centering
\begin{tabular}{|r|r|r|r|r|r|r|r|r|r|}
\hline
$y$ & $T$ & $\frac{\lambda_-}{(g v(T))^2}$ & $\bar{S}$ & 
    $\ln\left(\frac{A}{T^4}\right)$ & $\ln\left(\frac{R}{T^3}\right)$ \\[1.2ex]
\hline
0.1 & 123.85 & -0.2 $10^{-3}$  & 593.7 & -89.~ & -672.~\\ 
0.2 & 123.84 & -0.8 $10^{-3}$  & 149.5 & -81.4 & -222.7\\
0.3 & 123.82 & -1.9 $10^{-3}$  & 65.5 & -52.0 & -111.1\\ 
0.4 & 123.79 & -3.6 $10^{-3}$  & 35.4 & -38.3 & -68.5\\
0.5 & 123.76 & -5.8 $10^{-3}$  & 21.0 & -31.0 & -47.9\\  
0.6 & 123.72 & -8.6 $10^{-3}$  & 12.8 & -26.8 & -36.5\\ 
0.7 & 123.67 & -11.2 $10^{-3}$  & 7.6 & -24.4 & -29.8\\ 
0.8 & 123.60 & -11.6 $10^{-3}$  & 4.0 & -23.1 & -26.0\\ 
0.9 & 123.52 & -7.9 $10^{-3}$  & 1.5 & -23.3 & -25.0\\
\hline
\end{tabular} \label{tab10}
\caption{Numerical results for $\tilde{m}_H = \tilde{m}_W$}
\end{table}

\newpage

\begin{table}[ht]
\centering
\begin{tabular}{|r|r|r|r|r|r|r|r|r|r|}
\hline
$\frac{\tilde{m}_H}{\tilde{m}_W}$ & $y_s$ &  $\bar{S}$ & 
    $\ln\left(\frac{A}{T^4}\right)$ & 
    $\ln\left(\frac{A}{T^4}\right)/\bar{S}$ \\[1.2ex]
\hline
1/2 & 0.42 & 112.9 & -33.4 & 0.295 \\
3/4 & 0.33 & 101.2 & -45.5 & 0.450 \\
1   & 0.26 & 85.5 & -61.6 & 0.720 \\
\hline
\end{tabular}
\caption{Relative onset-temperatures and corresponding values
         for different Higgs masses}
\end{table}

\begin{table}[hb]
\centering
\begin{tabular}{|r||r|r||r|r||r|r|}
\hline
 & \multicolumn{2}{|c||}{$\tilde{m}_H = \frac{1}{2} \tilde{m}_W$} &
   \multicolumn{2}{|c||}{$\tilde{m}_H = \frac{3}{4} \tilde{m}_W$} &
   \multicolumn{2}{|c|}{$\tilde{m}_H =  \tilde{m}_W$} \\
\hline \hline
$\frac{\sigma}{g v(T)^3}$ & 
   \multicolumn{2}{|c||}{0.0219} &
   \multicolumn{2}{|c||}{0.0276} &
   \multicolumn{2}{|c|}{0.0338} \\
\hline \hline
$y$ & $\bar{S}$ & $\bar{S}_{\rm TW}$ & 
      $\bar{S}$ & $\bar{S}_{\rm TW}$ &
      $\bar{S}$ & $\bar{S}_{\rm TW}$ \\
\hline \hline
0.1 & 2216.4 & 2210.9 & 1086.9 & 1079.8 & 593.7 & 587.7 \\
0.2 & 558.1  & 566.7  & 273.8  & 275.6  & 149.5 & 149.0 \\
0.3 & 244.3  & 258.5  & 112.0  & 125.2  & 65.5  & 67.2  \\
0.4 & 131.5  & 149.5  & 64.8   & 72.0   & 35.4  & 38.5  \\
0.5 & 77.3   & 98.4   & 38.2   & 47.3   & 21.0  & 25.1  \\ 
0.6 & 46.4   & 70.4   & 23.1   & 33.7   & 12.8  & 17.7  \\
0.7 & 26.8   & 53.4   & 13.5   & 25.5   & 7.6   & 13.3  \\
0.8 & 13.6   & 42.2   & 7.0    & 20.1   & 4.0   & 10.4  \\
0.9 & 4.82   & 34.6   & 2.6    & 16.4   & 1.5   & 8.7   \\
\hline
\end{tabular}
\caption{The effective action of the critical bubbles in comparison with the 
          thin-wall approximation values}
\end{table}

\newpage

\section*{Figures}

\begin{figure}[th]           
\epsfxsize13cm    
\leavevmode                   
\centering                    
\epsffile{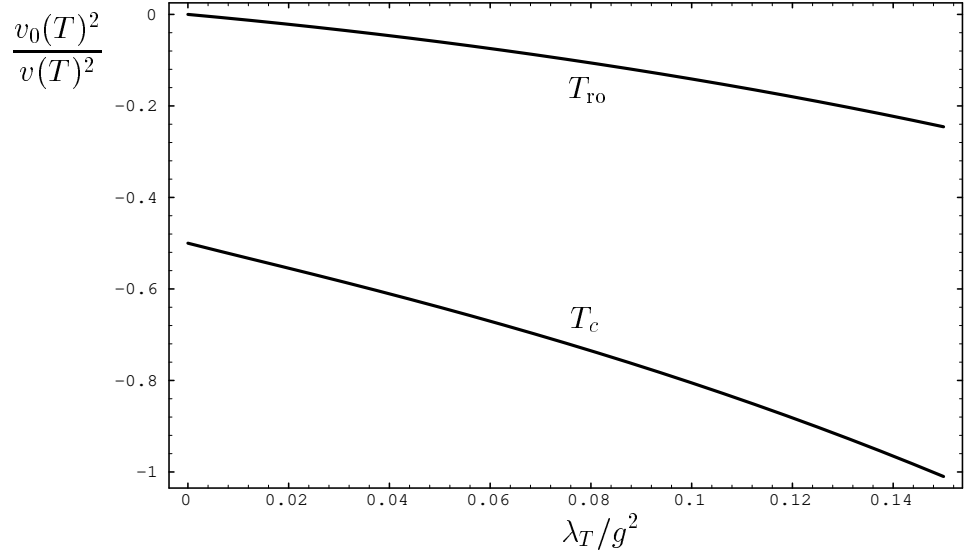}        
\caption{The parameter $\frac{v_0(T)^2}{v(T)^2}$ introduces the temperature
dependence into the masses and the potential. It is plotted versus 
$\frac{\lambda_T}{g^2}$ at the critical and at the roll-over temperature.
The plot range $0 \le \frac{\lambda_T}{g^2} \le 1.5$ covers the zero
temperature Higgs mass range $ 0 \le \tilde{m}_H \le \tilde{m}_W $.
}
\end{figure}

\vspace{1cm}

\begin{figure}[bh]           
\epsfxsize13cm    
\leavevmode                   
\centering                    
\epsffile{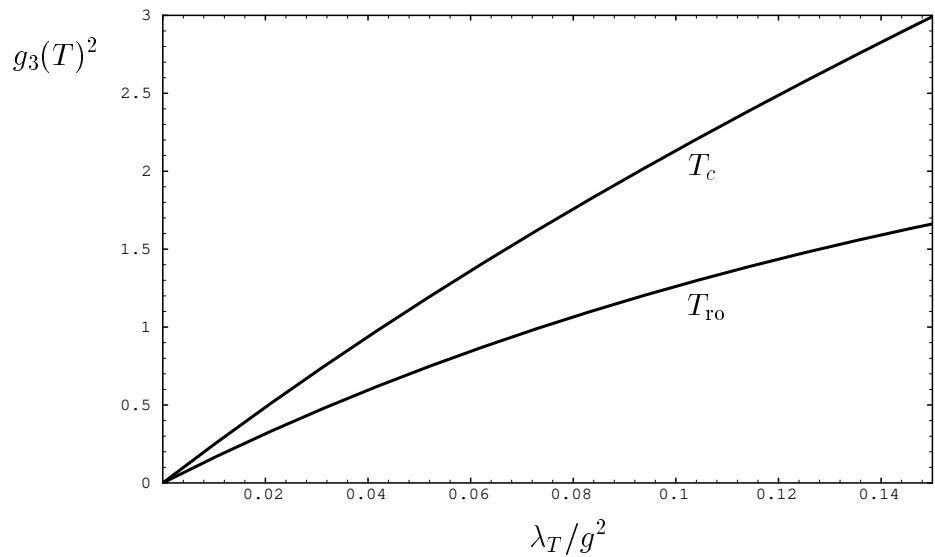}        
\caption{The effective 3-dimensional gauge coupling $g_3(T)^2$ versus
 $\frac{\lambda_T}{g^2}$ at the critical and at the roll-over temperature.}   
\end{figure}

\newpage

\begin{figure}[th]           
\epsfxsize13cm    
\leavevmode                   
\centering                    
\epsffile{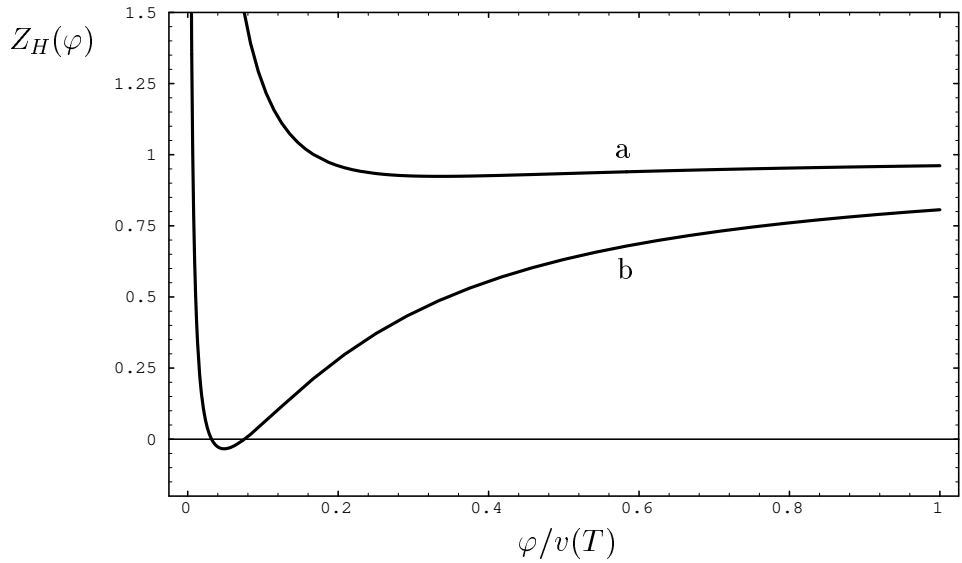}        
\caption{The $Z$-factor for the Higgs kinetic term $Z_H$ versus
         $\varphi$ in a) Landau and in b) 't~Hooft-Feynman gauge. 
         ($T=T_c$ and $\tilde{m}_H = \frac{1}{2}\tilde{m}_W$)}   
\end{figure}

\vspace{1cm}

\begin{figure}[bh]           
\epsfxsize13cm    
\leavevmode                   
\centering                    
\epsffile{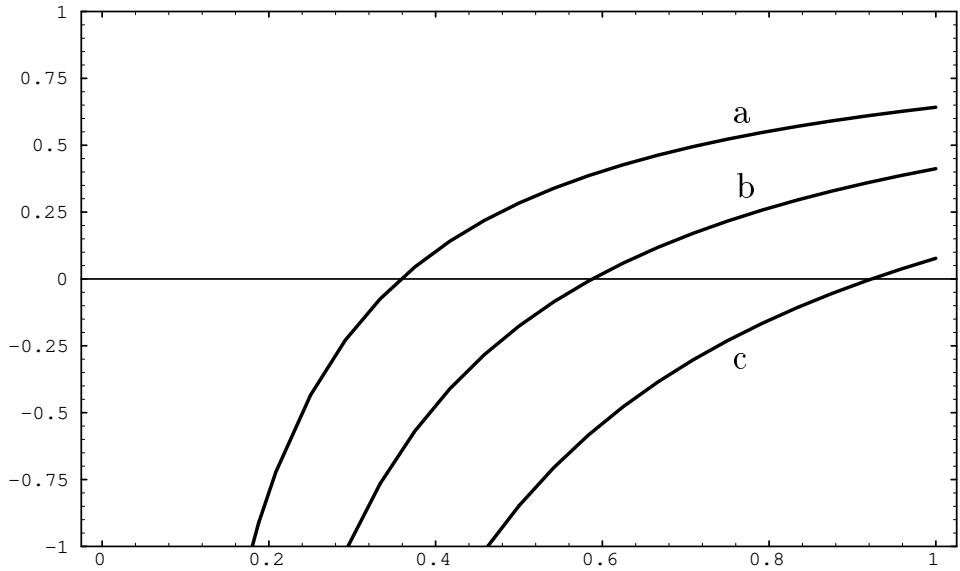}        
\caption{ $Z_{\rm gauge}$ at the roll-over temperature for
          $\tilde{m}_H =$  a) $\frac{1}{2}\tilde{m}_W$, 
   b) $\frac{3}{4}\tilde{m}_W$ and c) $\tilde{m}_W$.}   
\end{figure}

\newpage

\begin{figure}[th]           
\epsfxsize15cm    
\leavevmode                   
\centering                    
\epsffile{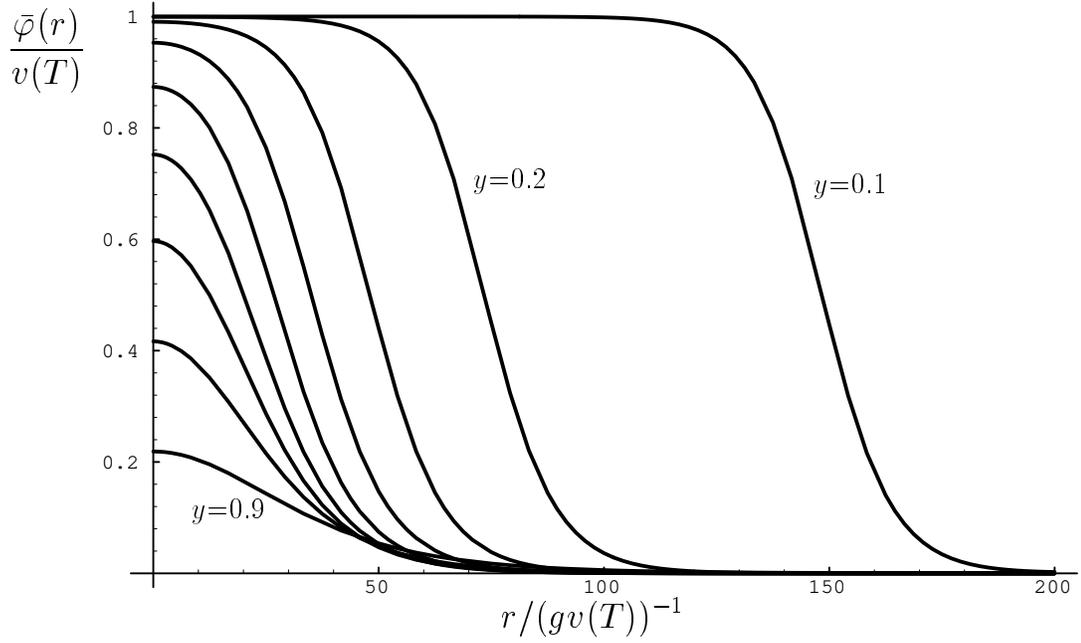}        
\caption{The critical bubbles are spherical symmetric. The profile functions
are plotted for nine temperatures which are defined via the parameter $y$
of eq.~(3.9). ($\tilde{m}_H = \frac{1}{2}\tilde{m}_W$)}   
\end{figure}

\vspace{1cm}

\begin{figure}[th]           
\epsfxsize13cm    
\leavevmode                   
\centering                    
\epsffile{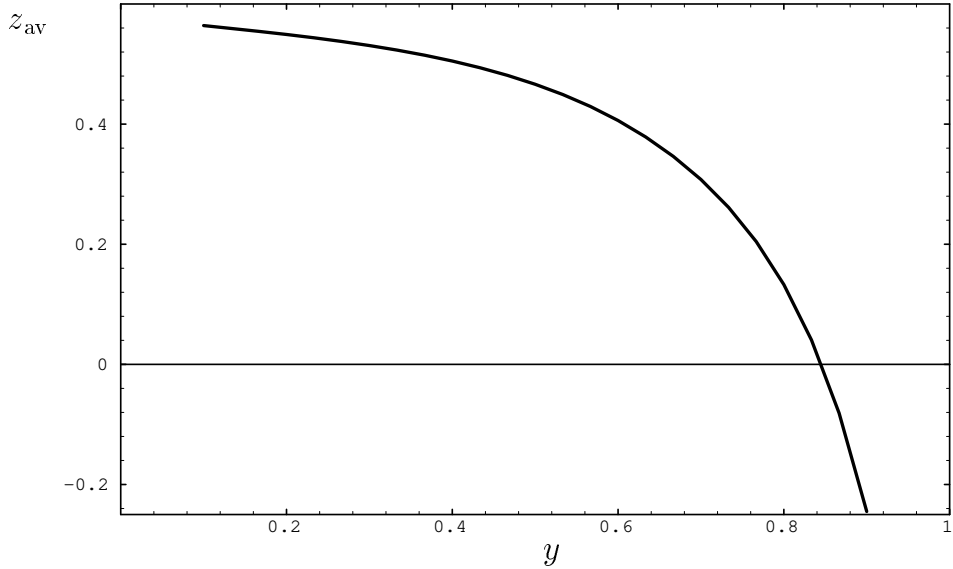}        
\caption{The average $z$ defined in eq.~(3.12) versus $y$. 
         ($\tilde{m}_H = \frac{1}{2}\tilde{m}_W$)}      
\end{figure}

\newpage

\begin{figure}[bh]           
\epsfxsize15cm    
\leavevmode                   
\centering                    
\epsffile{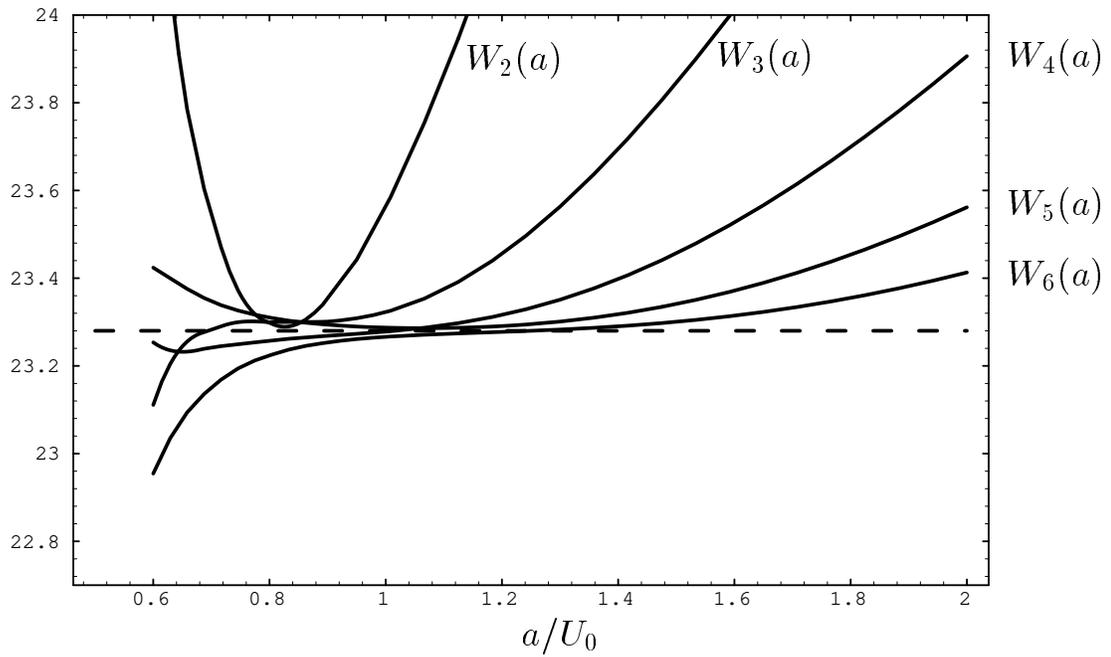}        
\caption{The functions $W_N(a)$ defined in equation (3.22) should 
converge towards a constant which is essentially the logarithm of the 
static prefactor. Here we give a typical plot of these functions.
The dashed line is assumed to be the limit.
($\tilde{m}_H = \frac{1}{2}\tilde{m}_W$ and $y=0.6$).
}   
\end{figure}

\newpage

\begin{figure}[ht]           
\epsfxsize14cm    
\leavevmode                   
\centering                    
\epsffile{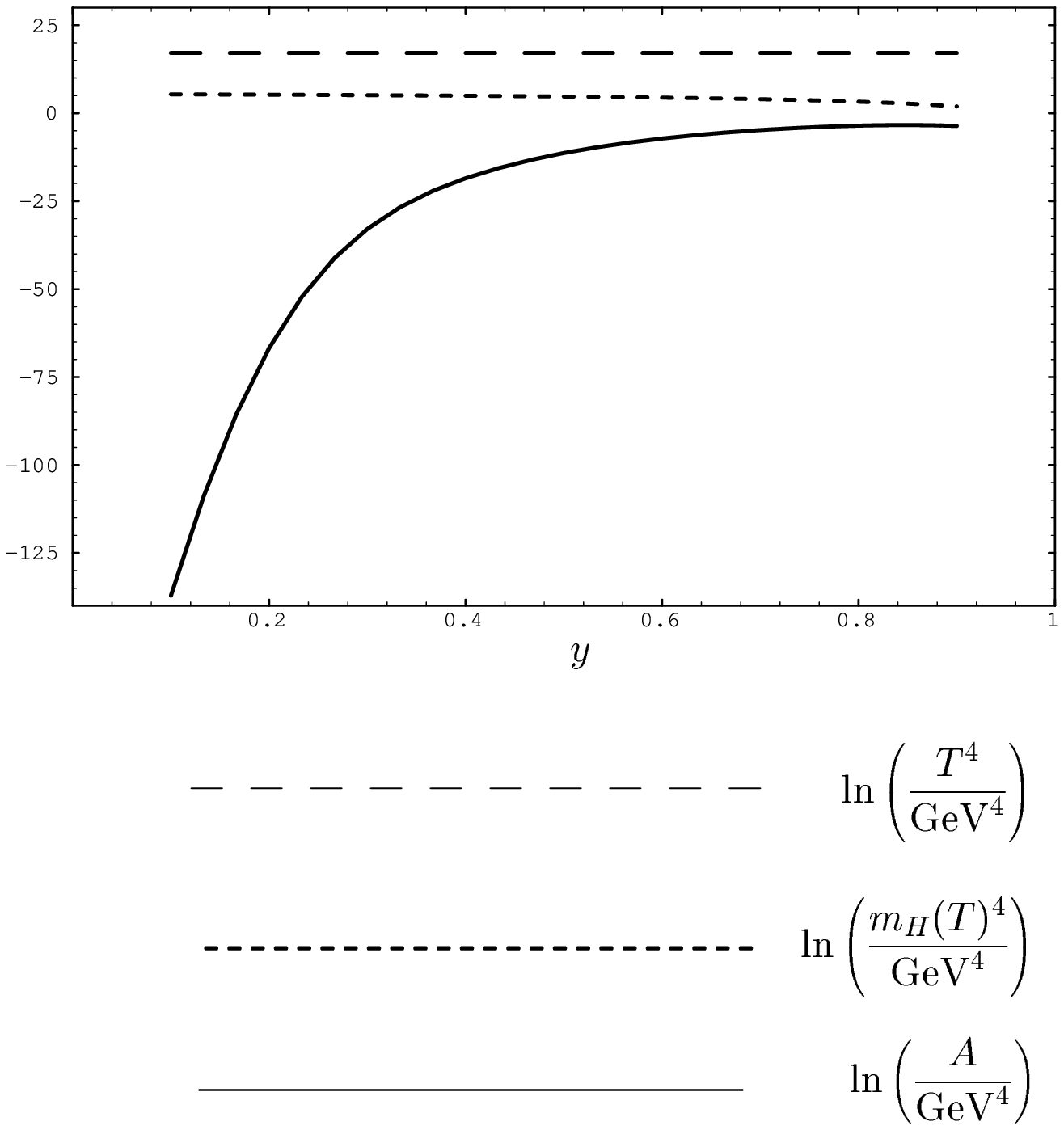}        
\caption{The static prefactor in comparison with dimensional estimates
         versus $y$. ($\tilde{m}_H = \frac{1}{2}\tilde{m}_W$).}   
\end{figure}

\newpage

\begin{figure}[ht]           
\epsfxsize14cm    
\leavevmode                   
\centering                    
\epsffile{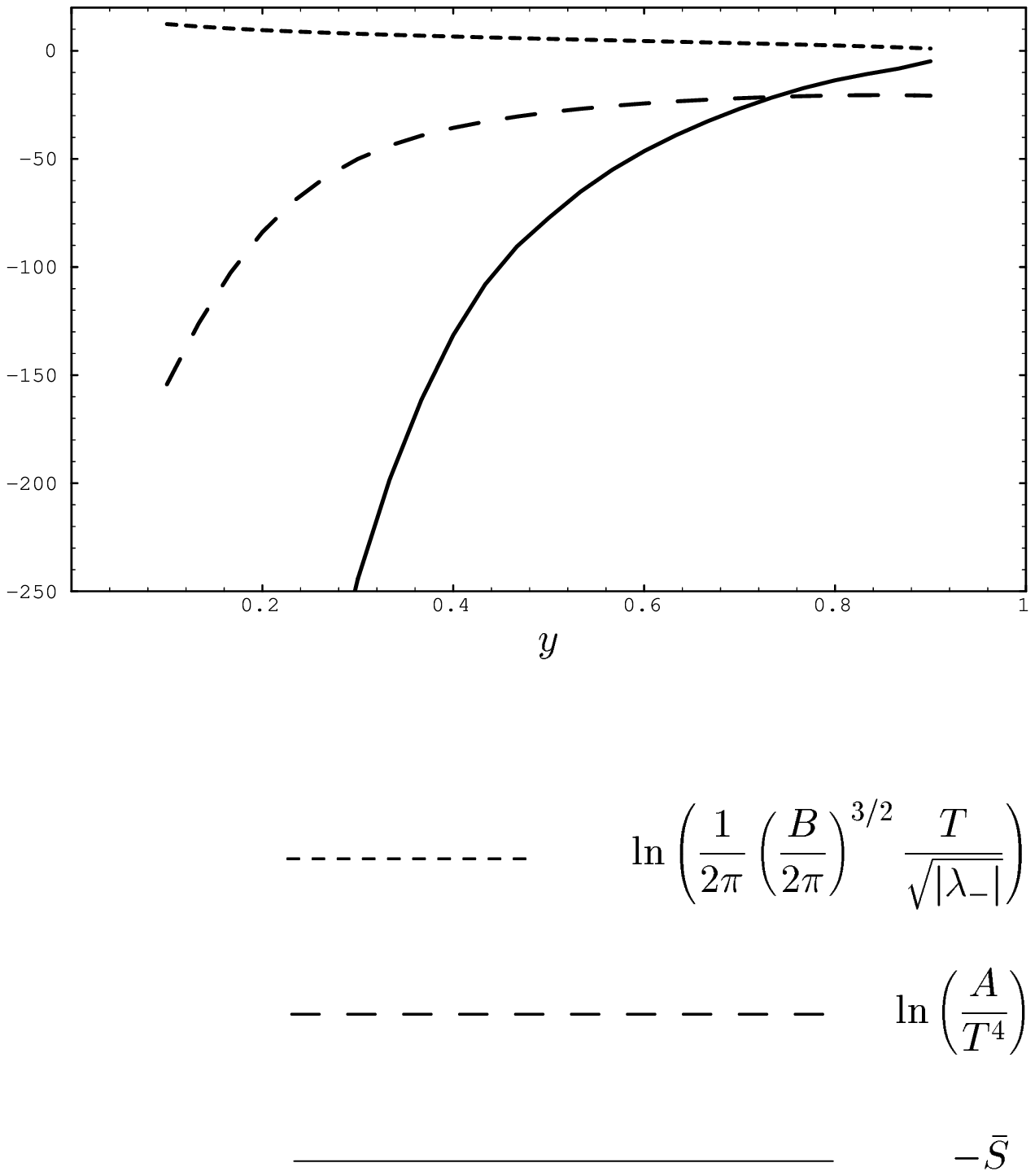}        
\caption{The summands of the logarithm of the nucleation rate (3.24).
   The main contributions are $-\bar{S}$ and $\ln\left(\frac{A}{T^4}\right)$.
   ($\tilde{m}_H = \frac{1}{2}\tilde{m}_W$)}   
\end{figure}

\newpage

\begin{figure}[ht]           
\epsfxsize14cm    
\leavevmode                   
\centering                    
\epsffile{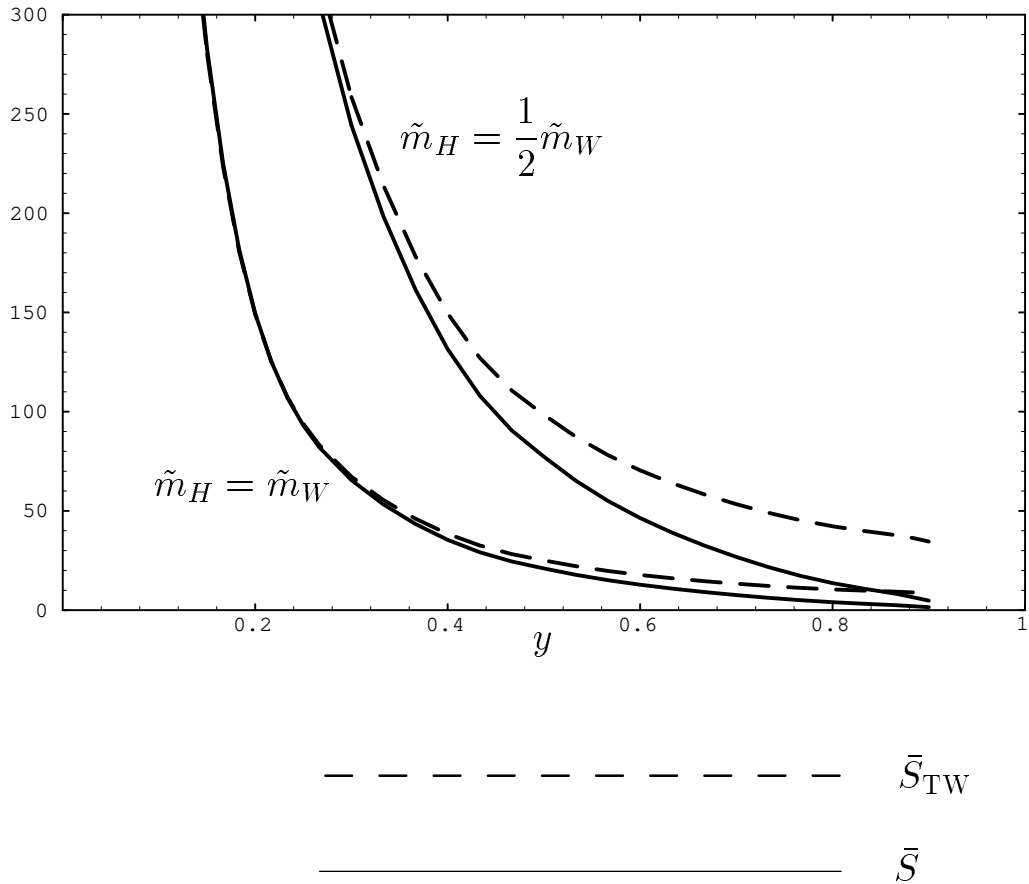}        
\caption{The effective action of the critical bubbles plotted in figure 5
in comparison with thin-wall estimates versus $y$.}   
\end{figure}

\end{document}